\documentclass[twocolumn]{aastex63}

\begin{document}
 
\newcommand{\kms}{km s$^{-1}\;$}
\newcommand{\msun}{M_{\odot}}
\newcommand{\rsun}{R_{\odot}}
\newcommand{\teff}{T_{\rm eff}}
\newcommand{\kep}{{\it Kepler}~}
\makeatletter
\newcommand{\Rmnum}[1]{\expandafter\@slowromancap\romannumeral #1@}
\newcommand{\rmnum}[1]{\romannumeral #1}
 
\title{Variability in the Massive Open Cluster NGC 1817 from K2: A
  Rich Population of Asteroseismic Red Clump, Eclipsing Binary,
  and Main Sequence Pulsating Stars
}
\shorttitle{Variability in Open Cluster NGC 1817}

\author[0000-0003-4070-4881]{Eric L. Sandquist}
\affiliation{San Diego State University, Department of Astronomy, San
  Diego, CA, 92182 USA}

\author[0000-0002-4879-3519]{Dennis Stello}
\affiliation{School of Physics, The University of New South Wales,
  Sydney NSW 2052, Australia}
\affiliation{Sydney Institute for Astronomy (SIfA), School of Physics,
  University of Sydney, NSW, 2006, Australia}
\affiliation{Stellar Astrophysics Centre, Department of Physics and
  Astronomy, Aarhus University, Ny Munkegade 120, DK-8000 Aarhus C,
  Denmark}

\author[0000-0002-4696-6041]{Torben Arentoft}
\affiliation{Stellar Astrophysics Centre, Department of Physics and
  Astronomy, Aarhus University, Ny Munkegade 120, DK-8000 Aarhus C,
  Denmark}

\author[0000-0003-2001-0276]{Karsten Brogaard}
\affiliation{Stellar Astrophysics Centre,
  Department of Physics and Astronomy, Aarhus University, Ny Munkegade
  120, DK-8000 Aarhus C, Denmark}
\affiliation{Institute of
  Theoretical Physics and Astronomy, Vilnius University, Sauletekio
  av. 3, 10257 Vilnius, Lithuania}

\author[0000-0002-8736-1639]{Frank Grundahl}
\affiliation{Stellar Astrophysics Centre, Department of Physics and
  Astronomy, Aarhus University, Ny Munkegade 120, DK-8000 Aarhus C,
  Denmark}

\author[0000-0001-7246-5438]{Andrew Vanderburg}
\altaffiliation{NASA Sagan Fellow}
\affiliation{Department of Astronomy, The University of Texas at Austin, Austin, TX 78712, USA}

\author[0000-0001-7017-678X]{Anne Hedlund}
\affiliation{San Diego State University, Department of Astronomy, San
  Diego, CA, 92182 USA}

\author{Ryan DeWitt}
\affiliation{San Diego State University, Department of Astronomy, San
  Diego, CA, 92182 USA}

\author{Taylor R. Ackerman}
\affiliation{San Diego State University, Department of Astronomy, San
  Diego, CA, 92182 USA}

\author{Miguel Aguilar}
\affiliation{San Diego State University, Department of Astronomy, San
  Diego, CA, 92182 USA}

\author{Andrew J. Buckner}
\affiliation{San Diego State University, Department of Astronomy, San
  Diego, CA, 92182 USA}

\author{Christian Juarez}
\affiliation{San Diego State University, Department of Astronomy, San
  Diego, CA, 92182 USA}

\author{Arturo J. Ortiz}
\affiliation{San Diego State University, Department of Astronomy, San
  Diego, CA, 92182 USA}

\author{David Richarte}
\affiliation{San Diego State University, Department of Astronomy, San
  Diego, CA, 92182 USA}

\author{Daniel I. Rivera}
\affiliation{San Diego State University, Department of Astronomy, San
  Diego, CA, 92182 USA}

\author{Levi Schlapfer}
\affiliation{San Diego State University, Department of Astronomy, San
  Diego, CA, 92182 USA}

\correspondingauthor{Eric L. Sandquist}
\email{esandquist@sdsu.edu}

\begin{abstract}
We present a survey of variable stars detected in K2 Campaign 13 within
the massive intermediate age ($\sim1$ Gyr) open cluster NGC 1817.
We identify a complete sample of 44 red clump stars in the cluster,
and have measured asteroseismic quantities ($\nu_{\rm max}$ and/or $\Delta
\nu$) for 29 of them.
Five stars showed suppressed dipole modes, and the
occurrence rates indicate that mode suppression is unaffected
by evolution through core helium burning.
A subset of the giants in NGC 1817 (and in the similarly aged cluster
NGC 6811) have $\nu_{\rm max}$ and $\Delta
\nu$ values at or near the maximum 
observed for core helium burning stars, indicating they have core
masses near the minimum for fully non-degenerate helium ignition.
Further asteroseismic study of these stars
can constrain the minimum helium core mass in red clump
stars and the physics that determines this limit.

Two giant stars show photometric variations on timescales similar to
previously measured spectroscopic orbits.  Thirteen systems in the
field show eclipses, but only five are probable cluster members.

We identify 32 $\delta$ Sct pulsators,
27 $\gamma$ Dor candidates,
and 7 hybrids
that are probable cluster members, with most new detections. We used
the ensemble properties of the $\delta$ Sct stars to identify stars
with possible radial pulsation modes. Among the oddities we have
uncovered are: an eccentric orbit for a short-period binary containing
a $\delta$ Sct pulsating star; a rare subgiant within the Hertzsprung
gap showing $\delta$ Sct pulsations; and two hot $\gamma$ Dor
pulsating star candidates.
\end{abstract}

\section{Introduction}

Galactic star clusters continue to be studied for insights into the
evolution of stars, and selecting a cluster for study based on its age
allows us to target stars of specific mass when they are in an
interesting stage of their evolution. One major reason for such
  targeting is that we can have the opportunity to examine the
  consequences of particular physics processes to gain a better
  theoretical understanding for modeling purposes. However, it is
  often the case that the observable characteristics of stars are most
  sensitive to the physics when the stars are evolving most rapidly,
  at the end of the main sequence and beyond. A consequence of this is
  that stars in a state of rapid evolution and short duration will
  necessarily be rare, and to maximize the likelihood of catching
  stars in such short stages we must study large (and if possible,
  uniform) samples of stars.

For these reasons, we present results from {\it Kepler} K2 Campaign 13
observations of the massive open cluster NGC 1817. NGC 1817 is about a
billion years old, meaning that stars of near $2 \msun$ are leaving
the main sequence and reaching the red giant clump. These stars
  turn out to be special in at least a couple of respects. First,
  these stars inhabit the main sequence portion of the pulsational
  instability strip, where $\delta$ Sct and $\gamma$ Dor variables are
  found. Because these stars typically excite many modes of
  oscillation, they promise to reveal information about the stellar
  interior by encoding it in the pulsation spectrum. Secondly, $2
  \msun$ is a transitional mass for stars, where they transition from
  non-degenerate ignition of core He in intermediate-mass stars to
  degenerate ignition in a He flash event for low-mass stars. The
  character of this transition is dependent on physics of the stellar
  core during H burning, including the details of core convection.

The cluster NGC 6811 in the main {\it Kepler} field has a nearly
identical age \citep{sand6811}, and there are other nearby clusters of
similar age that could also be studied to access stars near $2 \msun$.
We will discuss these other clusters as part of this paper, but the
major advantage of observing NGC 1817 is its large total mass. A
simple comparison of NGC 6811 with NGC 1817 reveals that NGC 1817 has
approximately five times as many stars in equivalent evolutionary
phases. This stellar sample provides the chance to examine the details
of the post-main sequence evolution with finer resolution.

The main drawback to the present study of NGC 1817 was that the
cluster was observed for only one K2 campaign of about 75 days, while
the main {\it Kepler} field (and NGC 6811) was observed for about 4
years. This limits the frequency resolution of our power spectra of
variable stars below, and prevents the detection of other variables
(like long-period eclipsing binaries or low-amplitude pulsators). In
addition, NGC 1817 has a higher reddening and a distance modulus that
is about a magnitude larger than that of NGC 6811. However, as we show
below, there was plenty to discover.

Before proceeding, we summarize some of the known characteristics
  of the cluster.  The reddening has been measured only a few times
with relatively large uncertainties, and better determinations would
be helpful.  \citet{hh77} measured the reddening of NGC 1817 using
$UBV$ two-color diagrams, and found $E(B-V) = 0.28\pm0.03$ for main
sequence stars, and $0.23\pm0.03$ for red giant stars.  \citet{bnphot}
presented Str\"{o}mgren photometry of the cluster, and estimated the
reddening to be $E(b-y)=0.19\pm0.05$ [$E(B-V) = 0.27\pm0.07$] from
main-sequence stars.

For chemical composition, \citet{overbeek} found a mean
[Fe/H]$=-0.05\pm0.02$, but this is probably systematically high based
on comparisons to other literature values.  \citet{jacob} found a mean
[Fe/H]$=-0.07\pm0.04$ from two giants, \citet{jacob11} found
[Fe/H]$=-0.16\pm0.03$ from 28 stars, while \citet{reddy} found a mean
[Fe/H]$=-0.11\pm0.05$ from 3 giants.  \citet{netopil} undertook a
study to homogenize previous literature abundance measurements, and
found $-0.11\pm0.03$ from 4 stars with high-quality spectroscopic
determinations and $-0.16\pm0.03$ from 28 stars with lower quality
spectroscopic determinations.  \citet{casamiquela} found [Fe/H] $=
-0.08\pm0.02$ and $-0.11\pm0.03$ from equivalent width and synthetic
spectrum analysis of 5 giant stars.  \citet{carrera} find [Fe/H] $=
-0.09\pm0.01$ for one giant star from the APOGEE project.

\section{Observational Material and Data Reduction}\label{obs}

\subsection{K2 Photometry}

NGC 1817 was observed during Campaign 13 of the K2 mission through
guest observer program GO 13053 (PI: Sandquist), although there were
overlaps with several exoplanet search programs. 924 of the brightest
cluster stars were identified from ground-based proper motion
membership information \citep{bnpm} and from color-magnitude diagram
position \citep{bnphot}. Most targets were observed with long cadence
(29.43 minute integration), with a handful of previously known
variables observed with short cadence (0.98 minute integration).

Because of the incomplete gyroscopic stabilization during the K2
mission, systematic effects on the light curve are substantial. The
primary source of the light curves we used in this investigation is
the K2SFF pipeline \citep{keplerav,k2av} that involves stationary
aperture photometry along with correction for correlations between the
telescope pointing and the measured flux. The K2SFF light curves typically
still had modest long-term trends, and for most of our scientific
purposes these trends were removed by fitting the points with a
low-order polynomial and subtracting the fit. When an eclipsing
binary was involved, only points outside of eclipse were fitted. For
giant stars with variations on the longest timescales, we divided by the
output of a sliding median filter covering 71 observations for most
stars. For 9 stars, the filter used a larger number of points (101,
121, or 161), with a single AGB star requiring a much larger number
(1001) in order to avoid erasing the photometric variations.

For the asteroseismic analysis of red clump stars below, we used
EVEREST light curves \citep{everest}. EVEREST uses pixel level
decorrelation to eliminate instrumental signals (ones that do not
recur) in pixels within the stellar aperture. We found that EVEREST
curves allowed us to reduce instrumental noise to a greater degree
than the K2SFF curves.

\subsection{Spectral Energy Distributions}\label{seddat}

For a small number of objects below, we compiled spectral energy
distributions (SEDs) from photometry in the literature and used
calibrations to put them on an absolute flux scale.  In producing the
SEDs, we collected photometry from the following sources: ultraviolet
photometry from the {\it Galaxy Evolution Explorer} ({\it GALEX};
\citealt{galex}), Str\"{o}mgren photometry from \citet{bnphot},
$BVI_C$ data from \citet{donati}, PSF magnitudes from Data Release 14
of the Sloan Digital Sky Survey (SDSS; \citealt{sdss}),
mean PSF magnitudes from the Pan-STARRS1 survey \citep{ps1desc},
the APASS survey \citep{apass}, Data Release 2 (DR2) of the Gaia archive
\citep{GaiaDR2}, Two-Micron All-Sky Survey (2MASS; \citealt{2mass})
photometry from the All-Sky Point Source Catalog, and from
the Wide Field Infrared Explorer (WISE; \citealt{wise}) survey.

\section{Analysis}\label{analy}

\subsection{Cluster Membership}

\citet{bnpm} presented proper motion membership determinations for
stars with $V < 15$, and \citet{kkpm} also studied the cluster's
proper motion membership as part of a larger survey of clusters.
However, DR2 from the {\it Gaia} spacecraft currently provides the
most precise proper motions and parallaxes to gauge cluster
membership. This is a particularly important point for the NGC 1817
cluster because the proper motions of its stars fall near the center
of the field star distribution, and proper motions alone are an
imperfect measure of membership.  \citet{cantat} have produced
membership probabilities for NGC 1817 and other clusters using {\it
  Gaia} DR2, and we have used their values for NGC 1817's proper
motion vector and average parallax. For the purposes of this paper, we
employed relatively simple criteria for membership: a star is a likely
member if its proper motion vector is within 0.35 mas yr$^{-1}$ of
$\mu_\alpha \cos \delta = 0.485$ mas yr$^{-1}$ and $\mu_\delta =
-0.890$ mas yr$^{-1}$, and a parallax within 0.2 mas of of
$\bar{\omega} = 0.551$ mas. The proper motion and parallax ranges
  were selected to extend well beyond the extremes of the distributions 
for star most obviously associated with the cluster.

For some of the stars (particularly giants), radial velocities provide
a strong additional constraint on cluster membership. \citet{m03},
\citet{jacob11}, and \citet{soubiran} measured radial velocities for
nearly all of the cluster giants. Surveys agree on a cluster mean velocity
near 66 km s$^{-1}$, with the largest survey of stars \citep{soubiran} finding
an average radial velocity of 66.02 \kms with a dispersion of 1.22 \kms from 40
stars.
We have eliminated stars with well-measured average velocities more
than 5 \kms from this mean value.

\subsection{Cluster Giants and Asteroseismic Analysis}\label{seis}

Star clusters with ages near 1 Gyr (like NGC 1817) produce red
clump stars through the non-degenerate ignition of helium in the cores
of the stars. These stars form the so-called ``secondary red clump''
(RC2), as opposed to the primary red clump (RC1) that is composed of
stars that undergo a helium flash when they ignite core helium in
degenerate cores of red giants. RC2 stars are often fainter than RC1
stars, due to a lower helium-burning luminosity that results from
  a lower-mass He core at ignition \citep{girardirev}, and minimums in
  luminosity and radius are expected at the transition between
  non-degenerate and degenerate ignition. Observationally, open
clusters of around 1.5 Gyr old show a combination of RC1 and RC2
stars, and the transitional mass is thought to be around $2.0-2.3
\msun$ \citep{girardi00}. RC2 stars are the progeny of less common,
  more massive stars than the giants in older clusters, and large
  uniform samples of those stars for study can be hard to come by.

Because NGC 1817 is a much more massive cluster than NGC 6811 (a
cluster of similar age in the original \kep field), it offers the
possibility of examining a much larger, uniform sample of clump stars
with masses greater than $2\msun$. A total of 8 clump stars were
identified as members of NGC 6811 \citep{sand6811}, with little chance
of any additional discoveries. All 8 of these stars were observed
during the main \kep mission and their asteroseismic oscillation
patterns were discussed by \citet{arentoft6811}. In NGC 1817, we
identified 44 likely giant members with 9 of these not observed during
the K2 campaign.  The sample in NGC 1817 is presented in Table
\ref{gianttab}.  Candidate members were identified using {\it Gaia}
parallaxes and proper motions. {\it Gaia} radial velocities were also
used for membership identification, but were not available for every
star. In other cases (mainly spectroscopic binaries), measurements
from \citet{m03,m07} and \citet{jacob11} were used to identify the
mean radial velocity.  We believe this sample for NGC 1817 is complete
or nearly complete based on a search out to about $2\degr$ from the
cluster center.

\begin{figure*}
\includegraphics[scale=0.8]{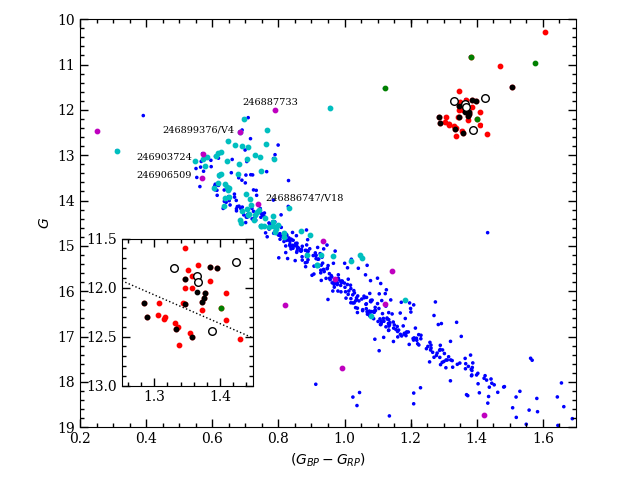}
\caption{{\it Gaia} color-magnitude diagram of NGC 1817, with giants
  (red), giants with asteroseismic measurements (black) and suppressed
  dipole modes (black circles), giants with single-lined spectroscopic
  binary orbits (green), pulsating stars (cyan), and eclipsing
  binaries (magenta) shown. EPIC IDs are given for selected eclipsing
  binaries. {\it Inset:} Zoom on the red clump, with a dotted line
  indicating the apparent gap between faint and bright clump stars.
\label{pulscmd}}
\end{figure*}

\begin{figure}
\epsscale{1.3}
\plotone{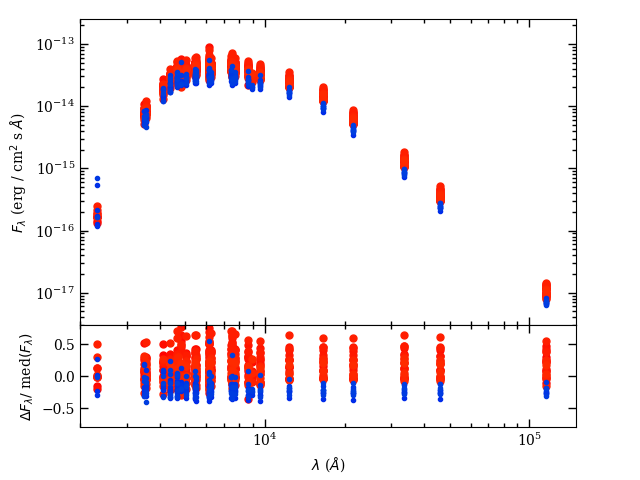}
\caption{{\it Top panel:} Photometric spectral energy distributions
  for stars identified as RC2 (red points) or fRC2 (blue points).
  {\it Bottom panel:} Fractional difference in flux for each star
  compared to the median for each filter, in order to flatten the
  SEDs. \label{giantseds}}
\end{figure}

An examination of the high-precision {\it Gaia} color-magnitude
diagram (Figure \ref{pulscmd}) gives an impression that a
  fainter subpopulation of the secondary red clump might be
  distinguishable, with a division at $G \approx 12.3$. Because
  of the potential importance of core He burning stars near the
  transition to degenerate ignition, we have labeled the
fainter stars in Table \ref{gianttab} as faint secondary clump stars
(fRC2). In Figure \ref{giantseds}, we show the photometric SEDs
  of all of the stars labeled RC2 or fRC2 to emphasize that the two
  groups are more clearly separated in magnitude within infrared bands
  (especially for $\lambda > 1 \mu$m), because the photometry is less
  sensitive to temperature differences there. Other properties of this
  group of stars will be discussed more below.

Asteroseismic studies have identified quantities that can be measured
in the oscillation frequencies of giant stars that connect to
fundamental properties of the stars. Most commonly, the frequency
separation of overtone modes $\Delta\nu$ and the frequency of maximum
power $\nu_{\rm max}$ reflect the average density \citep{kb95} and
surface gravity \citep{brownscale} of the stars, respectively, and
when algebraically combined, these return the mass and radius, with
some dependence on the effective temperature
\citep{kallinger}. Comparisons with the results of binary star mass
analyses has indicated there is a systematic overestimate of red giant
masses calculated from asteroseismic relations without corrections
(e.g. \citealt{bro16,bro18,gaulme}), although corrections have been
calculated to reduce the disagreements (e.g. \citealt{sharmacorr}).

A major benefit of cluster giant samples is that the ensemble of stars
has nearly identical age and chemical composition, and as a result the
stars should have similar masses. For RC2 stars though, there is generally a
larger variation in stellar mass than there is for the first ascent giants
seen in older clusters.

\begin{figure}
\plotone{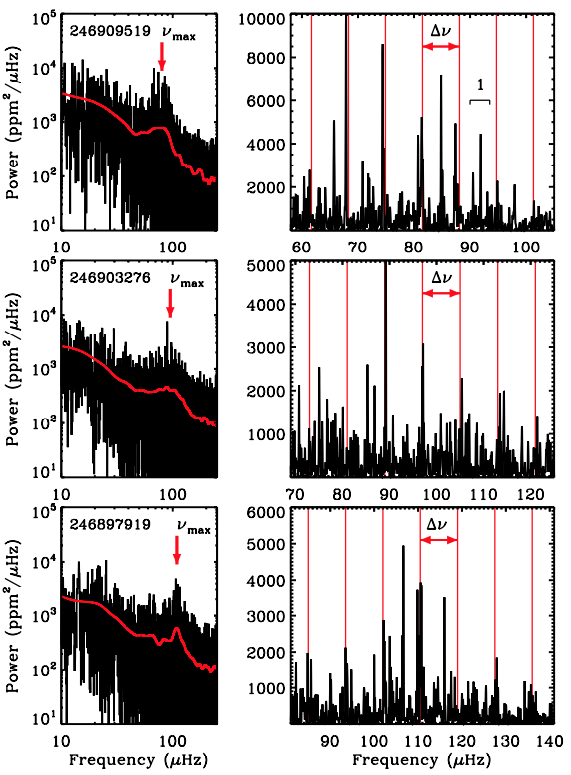}
\caption{{\it Left panels:} Power spectra of red clump star examples, with the smoothed spectrum shown in red, and the
  frequency of maximum power $\nu_{\rm max}$ indicated. {\it Right panels:}
  Power spectra zoomed on the solar-like excess power, with the large frequency
  separation $\Delta \nu$ indicated. In the top panel, an example of a detected $\ell = 1$ mode is shown.
  \label{specexamples}}
\end{figure}

We derived asteroseismic observables from the power spectra (Fourier
transforms of the K2 light curves) using two different pipelines. The
first followed the K2 processing procedure described in
\citet{stello15}, based on the method described in \citet{huber}. This
involved the use of a high-pass filter on the light curves to reduce
noise levels by decreasing spectral leakage of low-frequency power
into higher frequencies in the power spectrum, along with filling of
small data gaps (of no more than three sequential points) in order to
remove frequency peaks related to satellite repositioning. Our targets
in NGC 1817 are all of moderate luminosity compared to giants in older
star clusters, and because the peak power occurs at relatively high
frequency, so there are no issues with the high-pass filter's cutoff
frequency. The $\nu_{\rm max}$ frequencies we report are the central
values of a Gaussian envelope fit to the power spectrum of the
solar-like oscillations. The large frequency spacings $\Delta
  \nu$ are derived from autocorrelation of the power spectrum,
  followed by a Gaussian function fit to the one of the five highest
  peaks that is closest to the value expected from correlations
  between $\nu_{\rm max}$ and $\Delta \nu$. Three examples of power
spectra and the identified asteroseismic parameters are shown in
Figure \ref{specexamples}.
The second pipeline used the KASOC filter \citep{kasocfilt} to prepare
the light curves for asteroseismic analysis, followed by a
two-component fit to the power spectrum (with the asteroseismic
contribution again fitted with a Gaussian envelope;
\citealt{hand17}). This second pipeline was mainly used for validation
purposes, and when there was substantial disagreement between the
pipelines, we rejected stars from our final sample as unreliably
measured.  Because of the shorter period of observation in the K2
campaign, the frequency resolution is somewhat degraded relative to
the main \kep field. In addition, the NGC 1817 stars are somewhat fainter
than the analogous stars in NGC 6811 in the {\it Kepler} field.
As a result, only 29 of the stars had measurable $\nu_{\rm max}$ values,
and $\Delta \nu$ could be reliably measured for 20 of the stars.
This is still larger than the entire sample for NGC 6811, however.
On the other hand, \citet{arentoft6811} was able to measure the period
spacing $\Delta P$ for mixed modes for four of the NGC 6811 stars,
which was not possible in the short observations of NGC 1817.
There is only one other cluster of similar age where asteroseismology
has been done, and that is NGC 6633, where three stars were observed
using the {\it CoRoT} spacecraft cluster by \citet{n6633}.


\begin{longrotatetable}
\begin{deluxetable*}{crrrlccccclcccc}
\tablewidth{0pc} \tabletypesize{\scriptsize}
\label{gianttab}
\tablecaption{NGC 1817 Giants and Asteroseismology}
\tablehead{\colhead{EPIC ID} & \colhead{WEBDA} & \colhead{BN04\tablenotemark{a}} &
  \colhead{M\tablenotemark{a}} & \colhead{$G$} & \colhead{$(G_{BP}-G_{RP})$} & \colhead{Class} &
  \multicolumn{3}{c}{Membership} & \colhead{Notes\tablenotemark{b}} & \colhead{$\nu_{\rm max}$} & \colhead{$\sigma(\nu_{\rm max})$} & \colhead{$\Delta \nu$} & \colhead{$\sigma(\Delta \nu)$}\\
 & & & & & & & \colhead{RV} & \colhead{$\mu$} & \colhead{$\bar{\omega}$} & & \multicolumn{2}{c}{($\mu$Hz)} & \multicolumn{2}{c}{($\mu$Hz)} }
\startdata
246894008 & 1574 &     & 18 & 11.818 & 1.351 & RC2 & M & M & M & \\
246867463 & 1209 &  936 &    & 12.152 & 1.306 &fRC2 & M & M & M & Gaia \\
246868367 & 1208 &  935 & 57 & 12.003 & 1.347 & RC2?& M & M & M & SB\\
246868636 &  673 &  686 &  3 & 12.359 & 1.332 &fRC2 & M & M & M & \\
246871255 & 1456 &   26 &  1 & 11.027 & 1.471 & ?   & M & M & M & SB\\
246871689 & 1152 &  934 & 54 & 11.786 & 1.385 & RC2 & M & M & M &  & 57.91 & 0.95 & 5.25 & 0.06 \\
246872204 & 1459 &   27 &  9 & 11.592 & 1.347 & RC2 & M & M & M & \\
246876234 & 1408 & 4891 &  8 & 12.459 & 1.354 &fRC2 & M & M & M & \\
246879891 & 1412 & 4890 &  7 & 12.332 & 1.410 & RC2 & M & M & M & \\
246882666 &  286 &  158 &    & 12.325 & 1.315 &fRC2 & M & M & M & \\
246883940 &   73 & 7131 &C73 & 11.737 & 1.425 &AGB? & M & M & M & supp.& 40.07 & 0.92 & 4.24 & 0.03 \\
246884115 &   71 & 7132 &C71 & 12.586 & 1.337 &fRC2 & M & M & M &  &  \\
246884141 &   72 & 7133 &C72 & 11.801 & 1.330 &RGB? & M & M & M & SB;supp.& 61.65 & 1.64 & 6.35 & 0.08 \\
246885313 &  164 & 7751 &C164& 11.526 & 1.121 & ?   & M & M & M & SBO\\
246885840 &  212 &  152 &C212& 12.053 & 1.409 & RC2 & M & M & M &   & 51.8 &        & 5.65 & 0.50\\
246887834 &   40 & 7153 &C40 & 12.399 & 1.337 &fRC2 & M & M & M &   & 105.07 & 4.65 &     &      \\
246888025 &   81 & 7207 &C81 & 11.880 & 1.365 & RC2 & M & M & M & supp.& 62.54 & 1.35 & 5.70 & 0.07\\
246888508 &   79 & 7209 &C79 & 12.164 & 1.347 & RC2 & M & M & M &  & 81.19 & 1.76 & 7.03 & 0.15 \\
246888894 &   44 & 7156 &C44 & 10.830 & 1.381 & RC2?& M & M & M & SBO\\
246889824 & 1420 & 3669 & 13 & 12.275 & 1.305 &fRC2 & M & M & M & SB\\
246889908 &  211 &  149 &C211& 12.005 & 1.357 & RC2 & M & M & M &  & 66.47 & 1.93 &      &  \\
246890962 &   30 &  153 &C30 & 12.159 & 1.344 & RC2 & M & M & M &  & 78.39 & 1.60 &      &      \\
246891310 & 1135 &  522 & 47 & 12.145 & 1.373 & RC2 & M & M & M &  & 82.68 & 1.33 & 6.78 & 0.17 \\
246892343 &  244 &  523 &C244& 12.209 & 1.401 & RC2 & M & M & M & SBO & 84.45 & \\
246892429 &    8 & 7081 & C8 & 11.772 & 1.367 & RC2 & M & M & M &     & 56.55 & 1.78 &   &      \\
246892465 &  127 &  148 &C127& 11.935 & 1.385 & RC2 & M & M & M &  & 64.52 & 2.10 &      &      \\
246893162 &  206 &  145 & 11 & 11.493 & 1.505 & RGB?& M & M & M &  & 28.40 & 1.20 & 3.51 & 0.12 \\
246894024 &   12 & 7200 &C12 & 12.298 & 1.288 &fRC2 & M & M & M & SB & 98.02 & 5.77 & 8.36 & 0.30 \\
246895092 &   22 &  151 &C22 & 12.043 & 1.365 & RC2 & M & M?& M & SB & 71.91 & 2.32 & 6.65 & 0.87 \\
246896978 &   19 &  147 &C19 & 11.911 & 1.347 & RC2 & M & M & M & SB & 67.93 & 1.30 & 5.43 & 0.43 \\
246897011 & 1433 & 3667 & 15 & 11.878 & 1.358 & RC2 & M & M & M &    & 64.67 & 1.76 &      &      \\
246897327 &   64 &  146 &C64 & 11.797 & 1.396 & RC2 & M & M & M &    & 57.94 & 1.11 & 5.26 & 0.26  \\
246897919 &  121 &  161 &C121& 12.426 & 1.333 &fRC2 & M & M & M &    &102.82 & 2.75 & 7.86 & 0.76 \\
246899172 &  177 & 7361 &C177& 12.050 & 1.377 & RC2 & M & M & M &    & 74.41 & 2.42 & 6.55 & 0.25\\
246901772 &   56 & 7635 &C56 & 10.969 & 1.577 & RGB?& M & M?& M & SBO & 14.90 & 0.59\\
246903276 & 1114 &      & 48 & 12.439 & 1.388 &fRC2? & M & M & M & supp.& 95.00 & 3.45 & 7.97 & 0.14\\
246907326 &  185 & 7683 & 30 & 12.509 & 1.357 &fRC2 & M & M & M &    &100.33 & 5.82 & 8.50 & 0.19 \\
246907540 &  677 & 7686 & 20 & 12.231 & 1.372 & RC2 & M & M & M &    & 79.33 & 2.03 &      &     \\
246909519 & 1117 & 7429 & 42 & 12.102 & 1.375 & RC2 & M & M & M &    & 76.47 & 1.81 & 6.59 & 0.11  \\
246915802 & 1292 &      & 31 & 11.945 & 1.367 & RC2 & M & M & M & supp. & 67.56 & 1.10 & 5.92 & 0.06 \\
246923982 &      &      &    & 10.290 & 1.605 & AGB?& M & M & M & Gaia \\
246924226 & 1297 &      & 33 & 12.301 & 1.317 &fRC2?& M & M & M & Gaia & 81.9 &  & 8.07 & 0.59\\
246939662 & 1265 &      & 34 & 12.157 & 1.285 &fRC2?& M & M & M &    & 87.88 & 2.82 & 7.20 & 0.65 \\
246984210 &      &      &    & 12.527 & 1.431 &fRC2?& M & M & M & Gaia & \\
\multicolumn{15}{c}{Marginal Nonmembers} \\
246852586 & 1502 & 1127 &    & 12.346 & 1.430 &     & M &NM &M? &   \\
246853701 & 1476 & 2953 &    &  9.971 & 2.266 &     &NM & M &M? &   \\
246865268 &  555 &  688 &    & 12.704 & 1.451 &     &NM & M & M &   \\
246887966 &      &      &    & 12.736 & 1.395 &     &NM & M & M &   & & & & \\
\enddata
\tablenotetext{a}{BN04: \citet{bnpm}. M: \citet{m03}}
\tablenotetext{b}{SB: spectroscopic binary. SBO: spectroscopic binary with orbit determination. Gaia: Candidate identified in {\it Gaia} Data Release 2. supp.: suppressed dipole modes.}
\end{deluxetable*}
\end{longrotatetable}


\begin{figure}
\epsscale{1.3}
\plotone{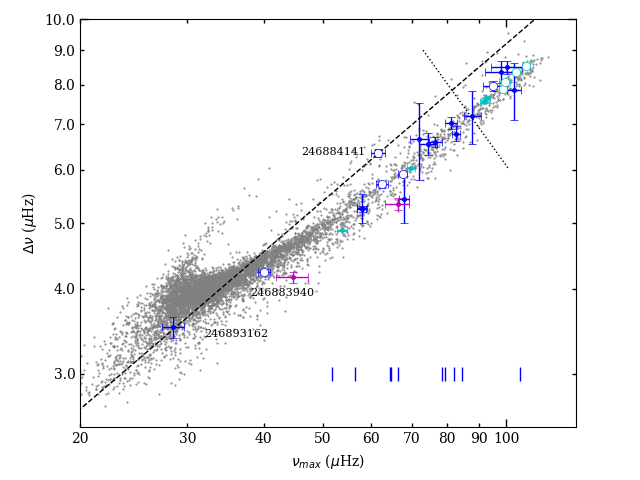}
\caption{Correlation of the asteroseismic measurables $\nu_{\rm max}$
  and $\Delta \nu$ for stars in NGC 1817 (dark blue), NGC 6633
  (magenta), and NGC 6811 (cyan), along with a dotted line separating
  the faint clump stars from brighter clump stars. Stars with suppressed
  dipole modes have open symbols, and stars having only $\nu_{\rm max}$
  measurements are shown with vertical line markers. Core helium
  burning stars from the {\it Kepler} sample of \citet{yu} are shown
  as gray points. The \citet{stello09} relation for solar-like
  oscillations is shown with a dashed line for reference.\label{asteropar}}
\end{figure}

A comparison of the asteroseismic $\nu_{\rm max}$ and $\Delta \nu$ for
the stars in the three clusters is shown in Figure
\ref{asteropar}. While the parameter correlations are consistent for
the clusters, different subgroups can be identified along the line.

\begin{itemize}
\item Four asteroseismically-measured stars in NGC 1817 appear to be
in evolutionary states other than the secondary clump based on their
low values.  Although EPIC 246883940 falls close to the bright end of
the red clump, its asteroseismic parameters indicate that it has
probably started to evolve toward the asymptotic giant branch
(AGB).
EPIC 246884141 has asteroseismic measures that are in the same range
as most RC2 stars, but when $\nu_{\rm max}$ and $\Delta \nu$ are
considered together the star falls on the correlation line for
first-ascent red giants \citep[e.g.][]{stello09}. The star is found
toward the blue side of the red clump, and in addition it is in a
spectroscopic binary and is observed to have suppressed dipole modes
(see \S \ref{suppressing}), so this star deserves more attention. The
asteroseismic data for EPIC 246893162 doesn't allow a clear
determination of its evolutionary state, but its red color may
indicate that it is a first-ascent red giant branch star (RGB) rather
than an AGB star.  Both EPIC 246884141 and 246893162 have unusually
low asteroseismic masses that we are unable to explain at this time.
EPIC 246901772 has the smallest $\nu_{\rm max}$ observed in NGC 1817
(and NGC 6811 as well).  Of the three measured stars in NGC 6633, two
are also probably evolved: HD 170174 is in the last stages of its core
helium burning, and HD 170053 appears to be an AGB star with very low
asteroseismic parameters (putting it off the edge of Figure
\ref{asteropar}).
\item The majority of the stars in NGC 1817 fall in an intermediate range
($\nu_{\rm max}$ between 50 and 85 $\mu$Hz, and $\Delta \nu$ between 4.5
and 7 $\mu$Hz) that corresponds to common secondary clump stars. One
of the measured NGC 6633 stars (HD 170231) is in this category of RC2
stars, as are the two brightest stars in NGC 6811.  In NGC 6811,
asteroseismic analysis of asymptotic period spacings of mixed modes in
four of the clump stars with the lowest $\Delta \nu$ values supports
the idea that the brightest stars are RC2 stars or are evolving toward
the asymptotic giant branch with some core helium remaining
\citep{arentoft6811}.
\item The last group is probably the most interesting, however. Six of
  eight stars measured in NGC 6811 are consistent with the four
  measured examples of the faint population of clump stars in NGC 1817
  (which we labelled ``fRC2''). The clearest asteroseismic faint clump
  stars are EPIC 246894024, 246897919, 246903276, and 246907326,
  although EPIC 246887834 has a $\nu_{\rm max}$ value that puts it
  among the fRC2 stars also. EPIC 246939662 is an additional
  borderline case. There is the appearance of a distinction between
  the bright and the faint RC2 stars in the two clusters, but there is
  not a clear division in the field star sample tabulated by
  \citet{yu}, as shown in Figure \ref{asteropar}.
\end{itemize}

In clusters, there is an observed lower limit to
  the luminosity of red clump stars, and the stars near this limit
  should define an upper limit for the asteroseismic parameters of
  clump stars as a consequence of higher average density.  According
to our observations and those of \citet{arentoft6811}, this appears to
be about $108 \mu$Hz for $\nu_{\rm max}$ and $8.5 \mu$Hz for $\Delta
\nu$. For a large sample of red clump stars identified with the help
of period spacings among mixed modes in {\it Kepler} samples, the
maximum observed value of $\Delta \nu$ was $8.73 \mu$Hz in
\citet{mosser14}, and $8.65 \mu$Hz in \citet{stello13}.  From the
larger \citet{yu} catalog of measurements from {\it Kepler} data, we
find maximum values of around $114-117$ $\mu$Hz in $\nu_{\rm max}$ and
around $8.9 \mu$Hz in $\Delta \nu$ after discounting a small number of
probable misidentified RGB stars (see Figure \ref{asteropar}).

\begin{figure*}
\plotone{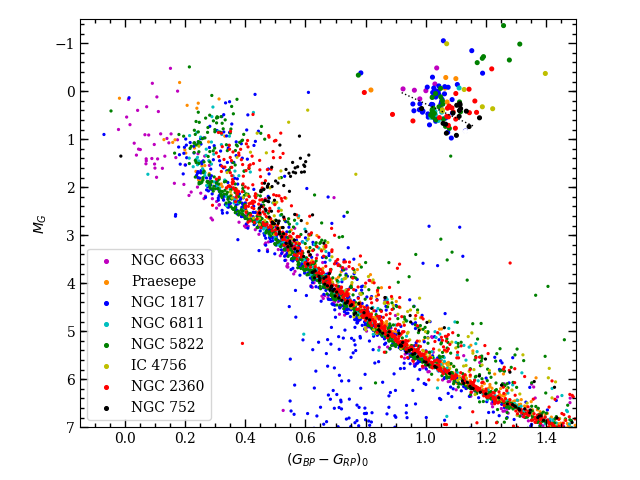}
\caption{{\it Gaia} dereddened color-absolute magnitude diagram for
  Milky Way open clusters with ages near 1 Gyr. \label{gyrcmd}}
\end{figure*}

\begin{figure*}
\plotone{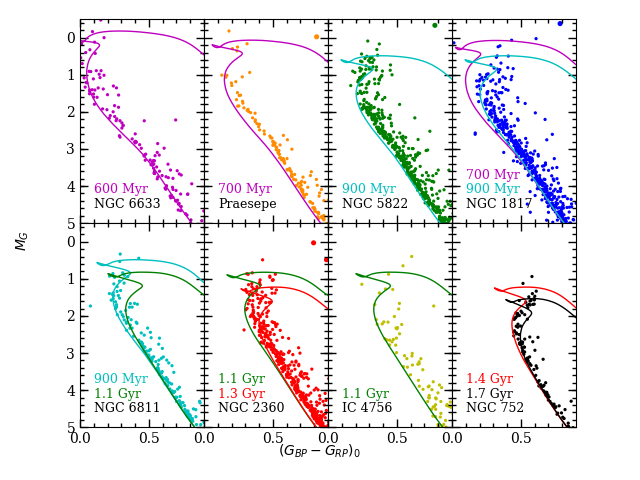}
\caption{{\it Gaia} dereddened color-magnitude diagrams, focused on
  the turnoff. MIST isochrones \citep{mist1} with convective
  overshooting and the labeled ages are shown. Isochrone [Fe/H] values are
  $-0.08$ for NGC 6633, NGC 5822, NGC 1817, and NGC 2360; $0.00$ for
  NGC 6811, IC 4756, and NGC 752; and $+0.06$ for Praesepe.
\label{tocmd}}
\end{figure*}

We can attempt to connect the available asteroseismic measurements
with the stellar properties in {\it Gaia} color-absolute magnitude
diagrams: both asteroseismic parameters are inversely proportional to
powers of the stellar radius, and radius is connected to CMD position.
In Figure \ref{gyrcmd}, we plot the {\it Gaia} CMDs of eight open
clusters with ages similar to NGC 1817 and well-determined mean
parallaxes. Important characteristics of the clusters are given in
Table \ref{gyrclusters}. Cluster members were selected using simple
cuts on proper motion and parallax, and the CMD was calculated using
mean parallax values from \citet{cantat} and reddening values from the
literature.  The main sequences align quite well, and the ordering of
the main sequence turnoffs gives an indication of the relative ages
(see Figure \ref{tocmd}): NGC 6633 and Praesepe appear slightly
younger than, and IC 4756 and NGC 2360 slightly older than the
intermediate group composed of NGC 1817, NGC 6811, and NGC 5822. NGC
752 is significantly older than the rest of the clusters. Most or all
of these clusters show significant color dispersion at the turnoff,
which complicates the relative age determination somewhat.

\begin{deluxetable*}{lrrrrrrrr}
\label{gyrclusters}
\tablewidth{0pt}
\tablecaption{Open Clusters with Ages near 1 Gyr}
\tablehead{ & \colhead{NGC 6633} & \colhead{Praesepe} & \colhead{NGC 1817} & \colhead{NGC 6811} & \colhead{NGC 5822} & \colhead{IC 4756} & \colhead{NGC 2360} & \colhead{NGC 752}}
\startdata
RA ($\deg$) & 276.845 & 130.10 & 78.139 & 294.340 & 226.051 & 279.649 & 109.445 & 29.223\\
DEC ($\deg$)&   6.615 & 19.67 & 16.696 &  46.378 &$-54.366$&  5.435  &$-15.632$ & 37.794\\
$\mu_\alpha \cos \delta$ (mas yr$^{-1}$) &  1.198 &$-36.047$&  0.485 &$-3.399$&$-7.422$&  1.260 & 0.385 &   9.810\\
$\mu_\delta$ (mas yr$^{-1}$)             &$-1.811$&$-12.917$&$-0.890$&$-8.812$&$-5.534$&$-4.927$& 5.589 &$-11.713$\\
$\bar{\omega}$ (mas) & 2.525 & 5.371 & 0.551 & 0.870 & 1.187 & 2.093 & 0.902 & 2.239\\
{[}Fe/H{]} & $-0.10$ & 0.12 & $-0.10$ & 0.02 & $-0.09$ & $-0.02$ & $-0.07$ & $-0.02$ \\
$E(B-V)$ & 0.19 & 0.027 & 0.23 & 0.07 & 0.15 & 0.13 & 0.09 & 0.044\\
\enddata \tablecomments{References for {[}Fe/H{]}: NGC 6633:
  \citep{jeff}; Praesepe: \citet{boesgaard}; NGC 1817: literature
  re-evaluation in introduction; NGC 6811: \citet{sand6811}; NGC 5822:
  \citet{ps18}; IC 4756: \citet{bagdonas}; NGC 2360, NGC 752:
  \citet{reddy}.  References for $E(B-V)$: NGC 6633: \citet{pena}; Praesepe:
  \citet{taylor06}; NGC 1817: \citet{hh77}; NGC 6811:
  \citet{sand6811}; NGC 5822: \citet{twarog}; IC 4756, NGC 2360: \citet{GaiaDR2CMD}; NGC 752:
  \citet{taylor07}.}
\end{deluxetable*}

\subsubsection{Model Comparisons}

Figure \ref{tocmd} shows the approximate matches between MIST
isochrones and the turnoffs of the eight clusters we are examining
here. Although we have plotted MIST models for comparison, BaSTI-IAC
models are very similar near the cluster turnoff. The purpose here is
simply to obtain a rough idea of the age from the turnoff in order to
see if age indications from the red clump are consistent. In a few
cases, we have plotted isochrones of other ages because they fit parts
of the red clump better.

\begin{figure}
\epsscale{1.3}
\plotone{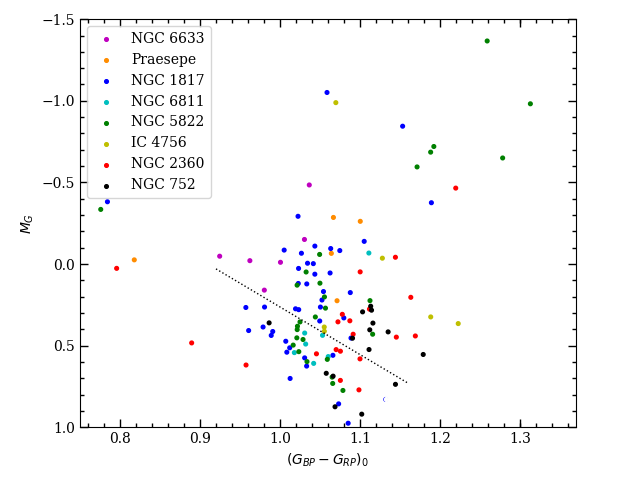}
\caption{Same as Figure \ref{gyrcmd}, but zoomed on the red clump. The
  dotted line shows our tentative dividing line between bright secondary
  red clump stars (RC2) and faint ones (fRC2).
\label{gyrclump}}
\end{figure}

\begin{figure*}
\plotone{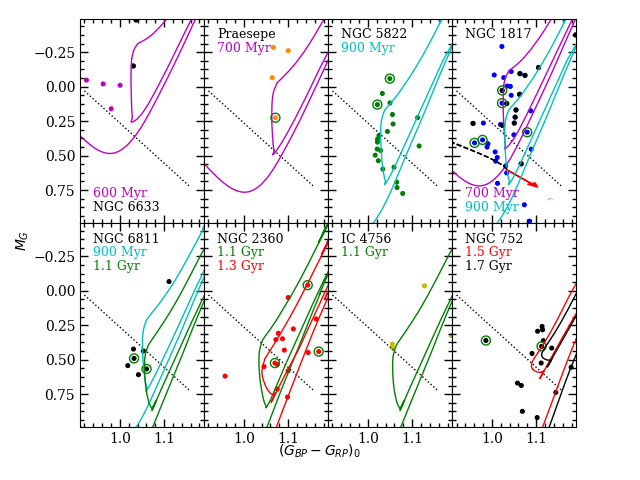}
\caption{{\it Gaia} dereddened color-magnitude diagrams, focused on
  the red clump. Black points in the NGC 6633, NGC 1817, and NGC 6811
  diagrams show stars with measured asteroseismic parameters.
  BaSTI-IAC isochrones \citep{basti} with convective overshooting and
  the labeled ages are shown, and [Fe/H] values are the same as in
  Fig. \ref{tocmd}. Open circles show stars with likely binaries with
  measured radial velocity variation. Dotted lines show the possible
  separator between bright and faint RC2 stars. In the NGC 1817 plot,
  a dashed line shows where photometric combinations of a faint clump
  star and a main sequence star would be, and a red arrow shows the
  reddening vector.
\label{clumpcmds}}
\end{figure*}

\begin{figure*}
\plotone{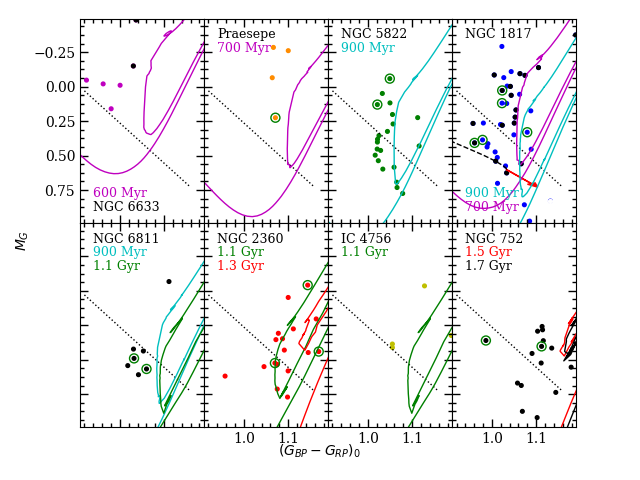}
\caption{{\it Gaia} dereddened color-magnitude diagrams, focused on
  the red clump, as in Figure \ref{clumpcmds}. MIST isochrones \citep{mist1} with 
  the labeled ages are shown. 
\label{clumpcmdsm}}
\end{figure*}

\begin{figure*}
\epsscale{1.3}
\plotone{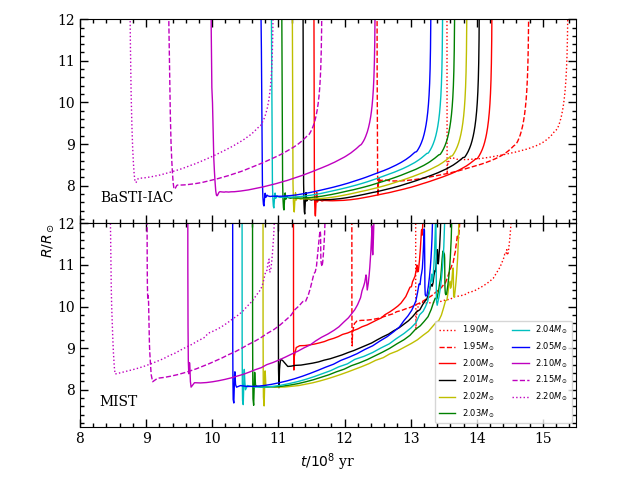}
\caption{Models of core He burning stars with core masses near the
  minimum from BaSTI-IAC \citep{basti} and MIST \citep{mist1}.
  Stable core He burning starts at the bottom
  of first steep section.
\label{rvst}}
\end{figure*}

Looking at the clump stars, five of the eight clusters shown (NGC
1817, NGC 6811, NGC 5822, NGC 2360, and NGC 752) appear to have stars
that fall in the faint secondary clump group (see Figures
\ref{gyrclump} - \ref{clumpcmdsm}). All of the faint clump stars that
also have asteroseismic measurements to date could have been
identified using $\nu_{\rm max}$ or $\Delta \nu$ as well. Models
\citep{girardi} indicate that these stars are likely to be made with
He core masses near the minimum allowable --- less massive stars
produce degenerate He cores that require substantially larger core
masses to ignite in He flash events, while more massive stars simply
generate larger convective cores on the main sequence due to more
powerful energy release. There is a similar change in He core burning
lifetime, with a maximum being reached when the He core mass hits its
minimum. This effect can allow minimum He core stars to be observed in
clusters at the faint end of the clump for a significant range of
  ages. Figure \ref{rvst} illustrates the effects on the radius and
  lifetime of core helium burning stars near the minimum core mass. To
  interpret the characteristics of the clump populations of these
  clusters and what they might say about the stellar interior physics,
  we will break the clusters into subgroups in the next subsection.

The youngest clusters in our sample (NGC 6633 and Praesepe, but the
Hyades could also be included), have ages in the range of $600-700$
Myr based on their main sequence turnoffs. For those ages, the clump
stars should all have ignited core helium burning in nondegenerate
gas. Models predict that the minimum luminosity (and
radius) for core He burning stars should be decreasing as the masses
of the stars reaching the clump are decreasing (see models on the left side of
Figure \ref{rvst}, or the synthetic CMDs in Figure 3 of \citealt{girardi00}).
So, these clusters are probably too young to produce
any of the faintest secondary clump stars. However, these clusters
only have eleven clump stars in total between them, so it is hard to say
that this is conclusive observational proof.
Of the stars in the three clusters,
only three giant stars in NGC 6633 have been studied
asteroseismically, and their masses place them comfortably in the
realm of nondegenerate core He ignition. The faintest 4 clump stars in
the cluster have not been analyzed however.
Four of Praesepe's giants have been observed with K2, and we predict that
their asteroseismic parameters will all fall comfortably within the
secondary clump.

The next older group of clusters (NGC 1817, NGC 6811, and NGC 5822)
have ages near 900 Myr using a turnoff-based age. Each of these
clusters have a substantial group of faint RC2 stars, but there is a
lot of variation in the percentage of clump stars the fRC2 stars
compose: from about 75\% in NGC 6811 down to 57\% for NGC 5822 to
about 25\% for NGC 1817.  The simple expectation from models is that
there should be a range of ages in which clump stars with core masses
near the minimum allowable are produced and persist. These clusters
are most likely ones in which secondary clump stars are being produced
in significant numbers but older clump stars (with more massive
non-degenerate cores) haven't died off yet. We will return to discuss
the distribution of the clump stars in the CMD below.

The oldest clusters (closer to 1.5 Gyr, like NGC 752 and NGC 7789;
\citealt{girardi00}) are probably the oldest systems that can produce
faint secondary clump stars. The faint RC2 stars would be among
  the longest residents of the clump, but some lower mass stars would
  have recently gone through a helium flash. It has previously been
shown that the distribution of stars in the CMD for clusters of this
age depends on the details of the transition between nondegenerate and
degenerate ignition of core He and the resulting rapid change in the
mass of the He core. A transition that occurs in a small range of
masses could allow significant numbers of stars of both kinds (RC1 and
RC2) to coexist in a cluster with significant differenes in radius
(and luminosity), as can be seen in the MIST models in Figure
\ref{rvst} for ages between about 1.1 and 1.3 Gyr. The BaSTI-IAC
models don't show a clear separation in the properties of the stars in
the same range of ages, and we believe this shows that the changes in
He core mass do not occur rapidly enough in these models.  Such a
scenario may still require effects (such as stellar rotation) that
introduce a spread in evolutionary outcomes for stars at a given mass
\citep{girardi00}.  Further characterization of the faintest secondary
clump stars may help address the issues of dispersion among the clump
stars, but also may reveal details of the convection that occurs
within them on the main sequence.

Returning to the CMD distribution of clump stars for the $\sim900$ Myr
old clusters, there are some problems in constructing a cohesive
explanation. NGC 5822 appears to be youngest according to the turnoff,
and the clump star positions are in crude agreement with the vertical
portion of the isochrone. The faint clump stars in NGC 6811 form a
tight group that can perhaps be qualitatively explained as resulting
almost exclusively from stars that had He core masses near the minimum possible.

The distribution of the clump stars in NGC 1817 is hard to reconcile
with models, however.  The large group of brighter clump stars are in
reasonable agreement with a young 700 Myr isochrone in terms of their
magnitude extent and colors, but an isochrone of that age doesn't fit
the turnoff. The superposition of bright and faint clump stars with
something resembling a gap between could imply a substantial drop in
radius over a small mass range, in analogy to what models predict for
older clusters like NGC 752. A clear gap is not seen in NGC 5822
though. The extension of the faint clump stars in color is also not
predicted by models. Unresolved binaries may be a contributor (see
Figure \ref{clumpcmds}), but would require a large fraction of the
faint clump stars to be affected. NGC 1817 also has the largest
reddening of any of the clusters we have discussed, so a contribution
to the color scatter by differential reddening is possible. The
reddening vector approximately parallels the extension of the faint
RC2 stars (again, see Figure \ref{clumpcmds}), although \citet{hh77}
and \citet{donati} did not find evidence of differential reddening in
the cluster from their ground-based studies.

\begin{figure}
\epsscale{1.3}
  \plotone{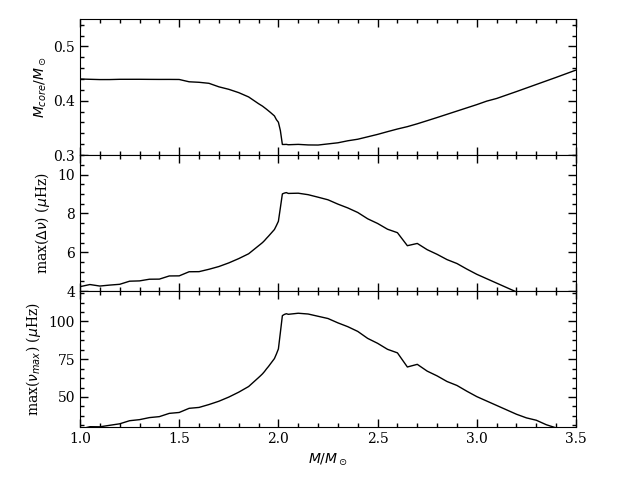}
\caption{Dependence of helium core mass and maximum asteroseismic
  parameters for helium core burning stars of differing total
  mass for MIST evolutionary tracks \citep{mist1}.
\label{clumpmod}}
\end{figure}

As a practical matter, the observation that there are maximum values
of the parameters $\nu_{\rm max}$ and $\Delta \nu$ in a vetted sample
of secondary clump stars can help to put limits on some of the
physics, and specifically effects on the minimum mass of the He core
in the red clump. Figure \ref{clumpmod} shows He core mass and maximum
values of the asteroseismic parameters during helium core fusion as a
function of total mass for MIST models \citep{mist1}. The maximum
asteroseismic measurements clearly correlate strongly with helium core
mass.  Overall, we find that there is fairly good agreement between
the observed maximum values of $\nu_{\rm max}$ and $\Delta \nu$ and
the MIST model predictions. The clump stars that have been
  measured using {\it Kepler} and {\it CoRoT} data in the field
  \citep{stello13,mosser14,yu} and in intermediate-age clusters
  comprise a large sample, and we can be reasonably assured that these
  stars sufficiently sample the faint secondary clump and its
  asteroseismic values.

\begin{figure}
\epsscale{1.2}
  \plotone{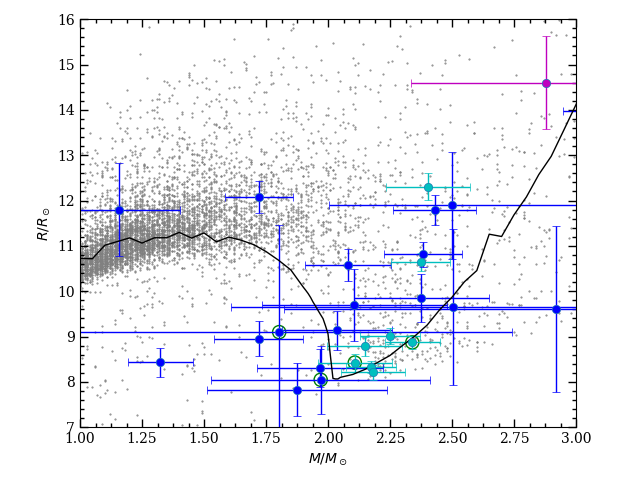}
  \caption{Asteroseismic mass and radius measurements from scaling
    relations for clusters NGC 1817 (blue), NGC 6811 (cyan), and NGC
    6633 (magenta), as well as field stars with {\it Kepler} data from
    \citealt{yu} (gray). Known binaries are identified with a green
    ring. Results for helium-burning stars with minimum radius from
    MIST models are shown with a solid line.\label{mvr}}
\end{figure}

The characterization of clump stars that have He cores near the
  minimum allowed mass can lead to a better understanding of the
  physics involved in the important transition between non-degenerate
  He ignition and degenerate He flash.  Models show that the
  amount of convective core overshoot during the main sequence
  strongly affects the stellar mass that produces a minimum-mass
  helium core, with stronger overshoot making the mass smaller
(e.g. \citealt{girardi00}). However, the overshoot does not strongly
affect what the mass of that helium core is (e.g. \citealt{sweigart};
Figure 12 of \citet{arentoft6811}). If the masses of these faint clump
stars can be determined precisely, the observations will put strong
constraints on the main sequence stars that later produce minimum-mass
helium cores. For the {\it Kepler} field-star sample, \citet{yu} used
the asteroseismic scaling relations with red clump star corrections to
$\Delta \nu$ \citep{sharmacorr} to calculate stellar masses and
radii. These are shown in Figure \ref{mvr}, and the radii show several
expected features like the sharp primary/secondary clump (RC1/RC2)
transition near $2.05\msun$ and the minimum radius among the RC2 stars
around $8\rsun$.  Typical uncertainties of around $0.16 \msun$ are
present in the individual field star measurements.

\begin{figure}
\epsscale{1.2}
  \plotone{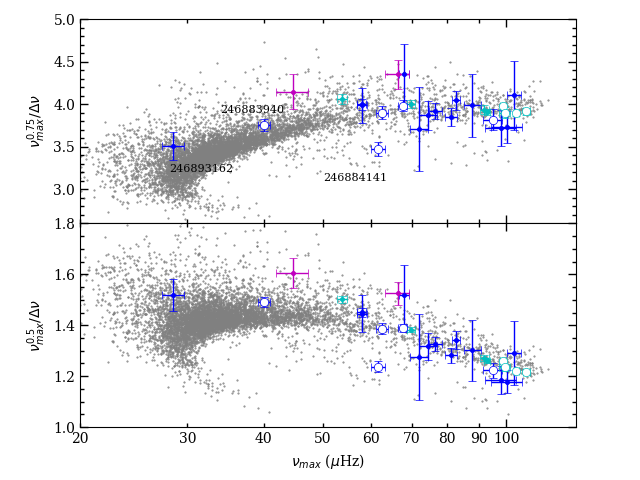}
  \caption{Mass-like (top panel) and radius-like (bottom panel)
    combinations of asteroseismic parameters for clusters NGC 1817
    (blue), NGC 6811 (cyan), and NGC 6633 (magenta), as well as field
    stars with {\it Kepler} data from \citealt{yu} (gray). \label{mrlike}}
\end{figure}

Following the same procedure for the cluster stars, we find that the
faintest NGC 6811 clump stars have masses between 2.11 and $2.33
\msun$, and the faintest NGC 1817 clump stars fall in the range 1.87
to $1.97 \msun$. Temperatures were computed from $(b-y)$ colors for
NGC 1817 stars, and were taken from \citealt{arentoft6811} for NGC
6811.  The NGC 6811 stars generally have smaller measurement
uncertainties, and have masses that agree well with the low mass edge
of the field star distribution.  As a check that the temperatures do
not change our main conclusions, we plot combinations of the
asteroseismic parameters that are proportional to mass and radius
separately (see Figure \ref{mrlike}) but do not correct for the minor
dependences on the temperature:
  \[ \left(\frac{\nu_{\rm max}}{\nu_{\rm max,\odot}}\right)^{3/4} \left(\frac{\Delta \nu}{\Delta \nu_\odot}\right)^{-1} = \left(\frac{M}{\msun}\right)^{1/4} \left(\frac{\teff}{T_{\rm eff,\odot}}\right)^{3/8} \]
  \[ \left(\frac{\nu_{\rm max}}{\nu_{\rm max,\odot}}\right)^{1/2} \left(\frac{\Delta \nu}{\Delta \nu_\odot}\right)^{-1} = \left(\frac{R}{\rsun}\right)^{1/2} \left(\frac{\teff}{T_{\rm eff,\odot}}\right)^{1/4} \]
The mass-like combinations for the NGC 1817 giants are consistent with
those of NGC 6811 with a couple of exceptions, and the faint RC2 stars
in the two clusters have fairly consistent radius-like combinations as
well.

To summarize, both the asteroseismic and photometric evidence suggests
that we can identify substantial populations of stars in $\sim 1$
Gyr-old open clusters that have He core masses at or near the minimum
possible, luminosities near a minimum, and
  average densities near a maximum.  Regardless of whether these faint
  red clump stars form a distinct class or not, models indicate that they are produced in a
  limited range of masses, and further study will help us learn about
  their internal physics. They are at a kind of phase transition in
stars, resulting from the change between non-degenerate and degenerate
He fusion ignition.  MIST models approximately match the mass of the
transition between primary and secondary red clumps and the rapidity
of the transition as a function of stellar mass, but higher precision
mass measurements would put stronger restrictions on the details of
convective core overshooting.  The models also predict a sharper rise
of radius versus mass among the RC2 stars than is seen in the field
stars. There are currently large statistical uncertainties and
  the possibility of significant systematic uncertainties in the
  asteroseismic masses though, and the systematic uncertainties in
  particular directly affect inferences about model physics like
  convective overshooting. (Overestimation of asteroseismic masses
  would lead to a need for greater overshooting in models.)  However,
further study of the lowest mass RC2 stars can realistically lead to
improved understanding of their cores.

\subsubsection{Suppressed Oscillation Modes}\label{suppressing}

Recently \citet{arentoft6811} found that the red clump stars of NGC 6811
could be split into two groups based on whether the stars had strong
dipole $l = 1$ modes. The stars without strong dipole modes are said
to have ``suppressed modes'', and in the case of the NGC 6811 sample,
the stars with suppressed modes were all found to be the faintest (and
possibly least evolved) of the clump stars.
\citet{cantiello} discussed the possibility that strong magnetic
fields in the cores of stars (including red clump stars) can affect
the propagation of mixed modes and prevent wave energy from escaping
the core. Dipole modes are especially susceptible to magnetic fields
due to strong coupling between the core ($g$ mode oscillation) and
envelope ($p$ mode oscillation). Their models assumed that the
magnetic fields were generated by a dynamo in the convective core of
the star during the main sequence, and the timescale for diffusion of
the field out of the core is far too long for significant
reduction. The field is thus expected to persist through the
first-ascent giant phase, but they admitted it was not clear what
would happen when convection started again during core helium burning.

Although we are unable to completely survey the stars in NGC 1817, we
were able to identify five stars with suppressed modes. We did this by
folding the power spectrum using the large frequency separation
$\Delta \nu$, and looking for peaks consistently spaced relative to
the $l = 0$ radial modes.
The power spectrum of one of the suppressed-power stars (EPIC
246903276) is depicted in Figure \ref{specexamples}.
The stars in NGC 1817 argue that suppressed modes are not restricted
to any one evolutionary state. One suppressed star is found among the
four faint red clump stars with asteroseismic measurements, three are
found among the ten bright RC2 stars with asteroseismic measurements,
and one is found among the two stars brighter than the clump that have
asteroseismology. These numbers are consistent with the results of
\citet{stellosupp}, who found that the occurrence rate of suppressed
modes among first-ascent giants of similar mass ($1.6-2.0 \msun$) was
at least 50\%. This appears to indicate that the mode suppression
continues unabated through helium burning in the core and the
convection that occurs there, or it appears the same fraction of the
time in stars in all of these evolutionary phases.

\subsubsection{Other Giant Variability}

Several of the giants are known to be single-lined spectroscopic binaries from
radial velocities, and four had orbits determined by \citet{m03} and
\citet{m07} with additional measurements in \citet{m08}. Two of
these had periods short enough that the K2 observations cover an
entire orbit (see Table \ref{specorb}). Although
none of the binary systems showed eclipses, two show signs of
variations that phase with or nearly phase with the orbit.

\begin{deluxetable*}{lccc}
\label{specorb}
\tablecaption{Orbital Elements of Giants in Spectroscopic Binaries}
\tablehead{\colhead{EPIC ID} & \colhead{246888894} & \colhead{246885313} & \colhead{246892343}\\
\colhead{WEBDA ID} & \colhead{44} & \colhead{164} & \colhead{244}}
\startdata
$P$ (d) & $68.0293\pm0.058$ & $165.760\pm0.021$ & $43.2485\pm0.0071$\\
$t$ - 2400000 & $49979.24\pm0.11$ & $49771\pm10$ & $45117.1\pm1.1$\\
$e$ & 0.000 & $0.020\pm0.007$ & 0.000\\
$\omega$ ($\deg$) & & $235\pm22$ & \\
$\gamma$ (km s$^{-1}$) & $65.69\pm0.16$ & $64.69\pm0.14$ & $65.79\pm0.13$\\
$K$ (km s$^{-1}$) & $21.00\pm0.22$ & $26.09\pm0.20$ & $4.61\pm0.18$\\
$N_{obs}$ & 24 & 26 & 11\\
Orbit Ref.\tablenotemark{a} & M03 & M03 & M07\\
\enddata
\tablenotetext{a}{M03: \citet{m03}. M07: \citet{m07}.}
\end{deluxetable*}

\begin{figure}
  \plotone{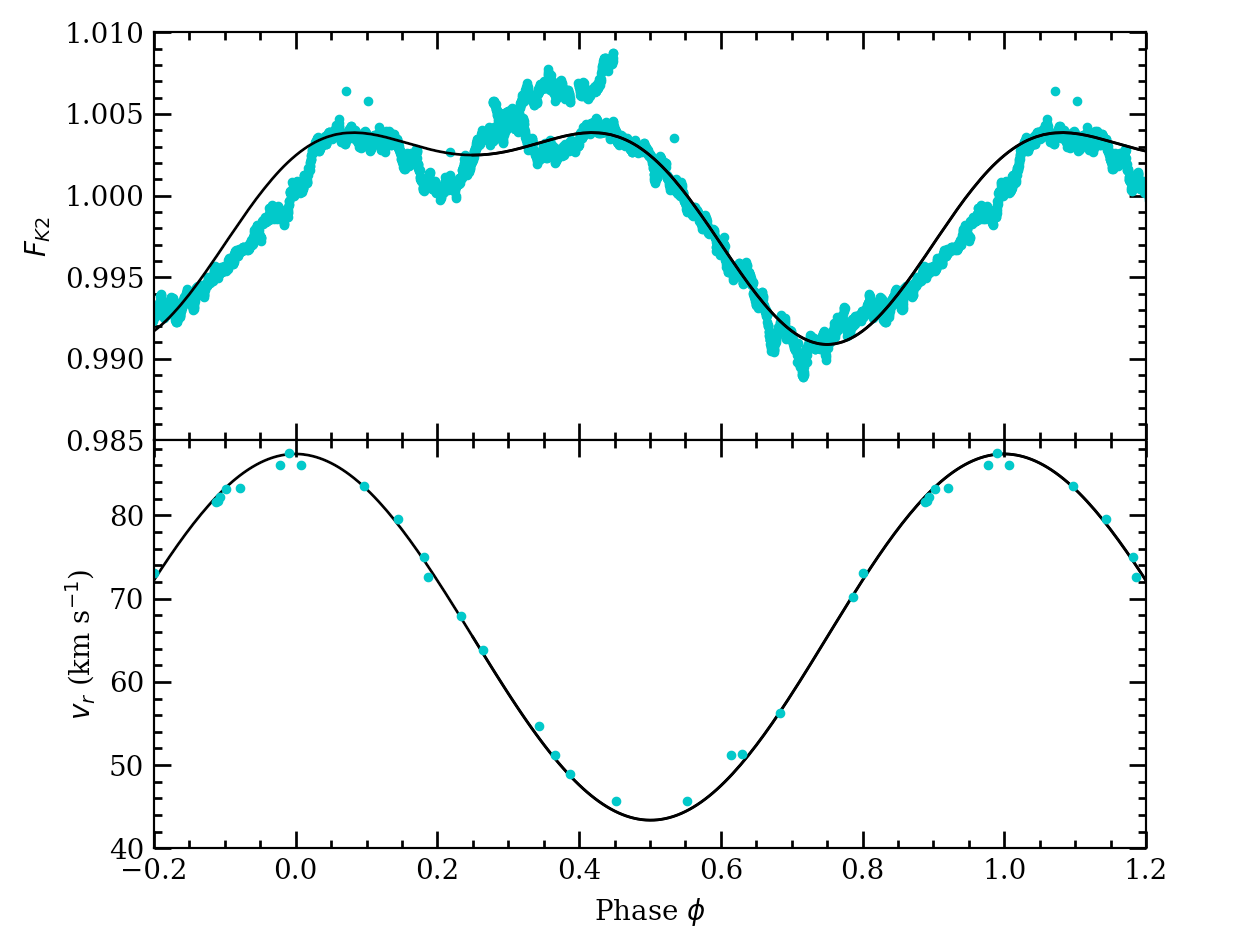}
\caption{K2 light curve ({\it top}) and radial velocity curve ({\it
    bottom}; \citealt{m03}) for the red clump star binary EPIC
  246888894, phased to the ephemeris of \citeauthor{m03}. Solid lines
  show a preliminary model fit to both curves.
\label{c44phase}}
\end{figure}

\begin{figure}
\epsscale{1.3}
\plotone{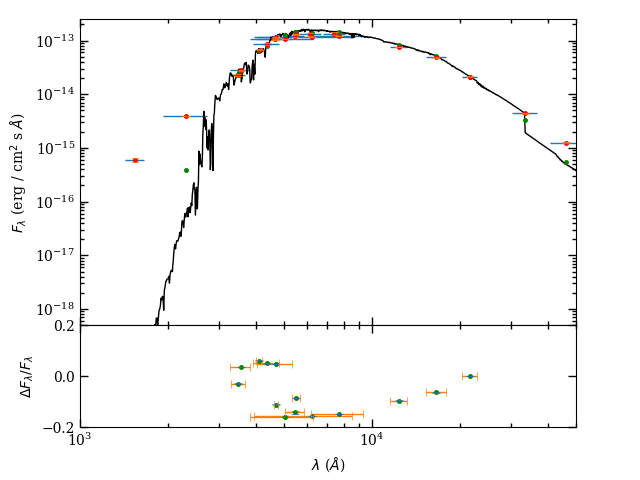}
\caption{{\it Top panel:} Spectral energy distribution for the star
  24688894 from photometry (orange points), compared with an ATLAS9
  model with $T_{\rm eff} = 5000$ K, $\log g = 3.0$, and $E(B-V) =
  0.23$ (with integrated fluxes in the filter bands showed with green
  points). Horizontal bars show the approximate wavelength range of
  each filter used. {\it Bottom panel:} Fractional flux residual
  between the observations and the model fit.
\label{c44phot}}
\end{figure}

For EPIC 246888894, the velocity variations of the giant ($K =
21.00\pm0.22$ km s$^{-1}$) are fairly substantial, but we see no signs
of an eclipse. We do, however, find photometric variations that phase
approximately with the radial velocity orbit ($P = 68.0293$ d; see
Figure \ref{c44phase}), with minimum light occurring when the
  brighter giant star was closest to us.  Qualitatively, this is
  consistent with the reflection effect, where the light of the
  companion star significantly illuminates the side of the giant that
  is facing it. This binary is one of the brightest cluster members,
  and it shows clear sign of the companion in its spectral energy
  distribution, particularly in the ultraviolet in both {\it GALEX}
  NUV and FUV bands (see Figure \ref{c44phot}). However, the light
  curve does not repeat particularly well during the phases of
  overlap, and this might indicate that the variation is at least
  partly due to spots.  Further analysis of the light curve will be
  postponed until additional photometric monitoring is done or the K2
  light curve can be detrended more reliably. But the indications are
  that the companion might be detectable in spectra, and that
  additional information on the system might be obtained from the
  photometric variations.

\begin{figure}
  \plotone{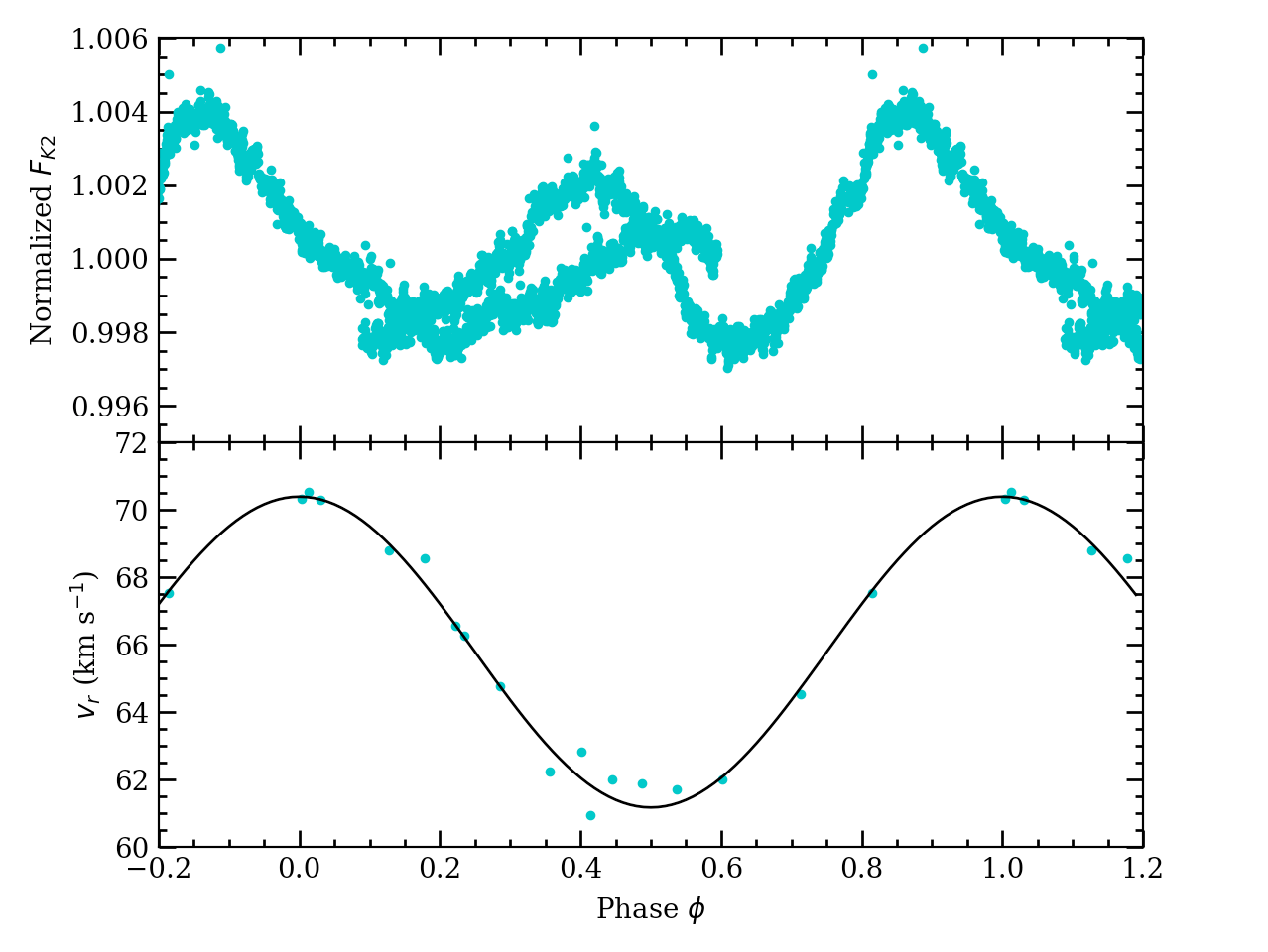}
\caption{K2 light curve ({\it top}) and radial velocity curve ({\it
    bottom}; \citealt{m03}) for the red clump star binary EPIC
  246892343, phased to the spectroscopic circular orbit fit (solid line) of \citet{m07}.
\label{c244phase}}
\end{figure}

EPIC 246892343 was identified as a spectroscopic binary by
\citet{m03}, but the orbital parameters were finally measured by
\citet{m07}. In this case, we used the K2 Pre-Search Data Conditioning
pipeline light curve, with a 20 point median filter used to smooth
some of the instrumental artifacts. The K2 light curve shows
photometric variations, and these have been phased to the
spectroscopic orbit ($P = 43.2485$ d) in Figure \ref{c244phase}. Where
the parts of the light curve overlap in orbital phase, there is only
mediocre agreement, and a photometric period near 53 days would
produce better agreement. It is not clear, however, how much the
detrending of the light curve affects this conclusion. Based on the
spectroscopic period, the two observed brightness maxima are separated
by roughly 0.5 in phase, and the light curve has its strongest minimum
at phase $\phi \approx 0.65$. Future space-based photometric
observations might help clarify what part of the light curve comes
from binary star effects like reflection and ellipsoidal variations
(e.g. \citealt{beer}), and what part comes from spots that are out of
synchronism with the orbit. We were able to detect solar-like
  oscillations for this binary as well, although only $\nu_{\rm max}$
  was measurable. Shorter period giant star binaries appear to have
  suppressed oscillations \citep[e.g.,]{gaulme}, so it is interesting
  that this clump star binary does show oscillations. The clump star
  would have already gone through its red giant phase with a maximum
  radius near $25\rsun$, which is still less than half the likely
  orbital separation with the companion object ($\sim66\rsun$). It is
  possible the clump star has not lost significant mass, so further
  seismic observation could still settle the star's mass and radius.

The period of a third giant star binary (EPIC 246885313) is more than
twice the duration of the K2 observations, so we could only get a
partial sampling of potential variations with timescales similar to the orbit period. The long-term
instrumental variations in K2 are consistent with those of other
nearby stars and large enough that we do not feel that there is a
reliable detection of variations that phase to the orbit.

\subsection{Eclipsing Binary Stars}\label{ecl}

\begin{figure*}
\plotone{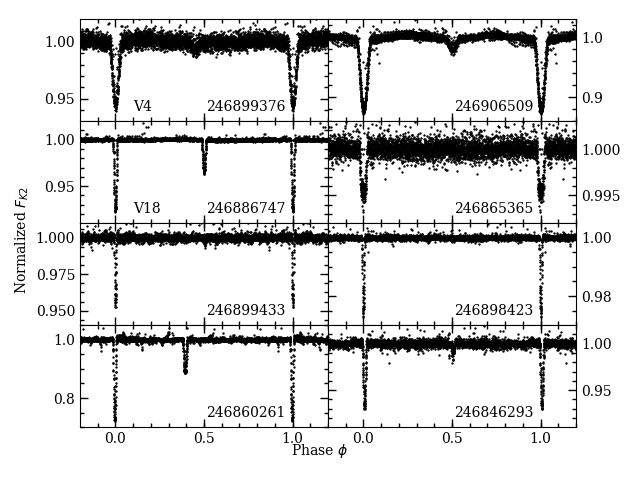}
\caption{Phased {\it K2} light curves for detached eclipsing binaries in the field of NGC 1817.
  \label{ebphase}}
\end{figure*}

\begin{figure*}
\plotone{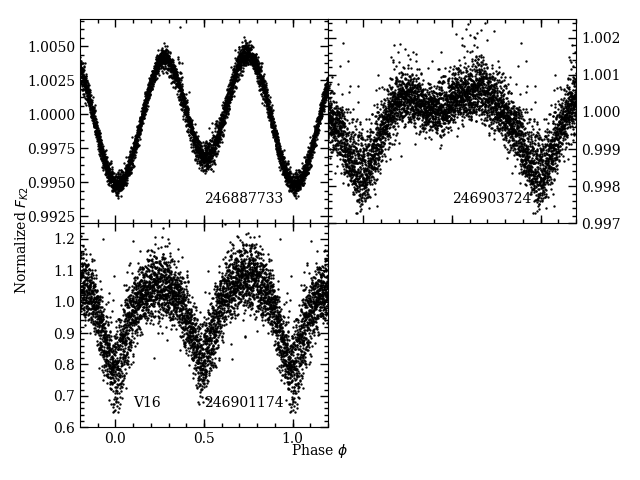}
\caption{Phased {\it K2} light curves for near-contact binaries in the field of NGC 1817.
\label{ewphase}}
\end{figure*}

\begin{figure}
\epsscale{1.3}
  \plotone{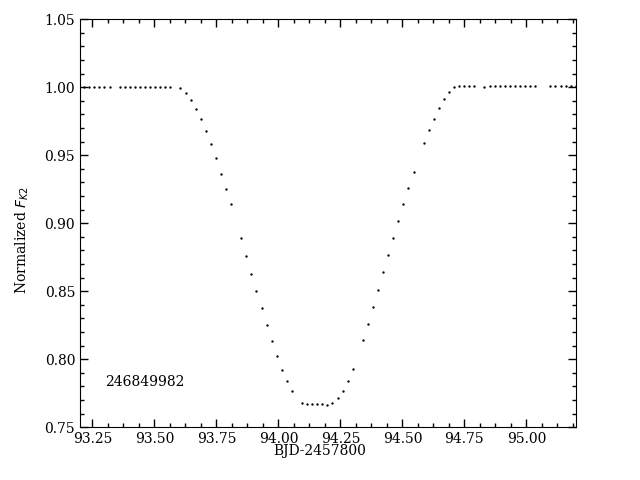}
\caption{{\it K2} light curve for EPIC 246849982 with a single detected eclipse.
\label{singlee}}
\end{figure}

Three eclipsing binary star members were discovered by
\citet{arentoft1817}, and the {\it K2} dataset has allowed us to
determine the ephemerides for all three stars (labeled V4, V16, and
V18) and produce complete light curves. In addition, we have
discovered 7 new detached eclipsing binaries (types EA and
EB\footnote{EA binaries have nearly spherical components, while EB
  star have somewhat ellipsoidal stars that results in out of eclipse
  variation.}) including two with periods longer than the {\it K2}
observing campaign, and three additional contact or near-contact
binaries (type EW). The data for these stars are given in Table
\ref{ecltab}, and the light curves are shown in Figures \ref{ebphase},
\ref{ewphase}, and \ref{singlee}.

\begin{deluxetable*}{lcrccclccl}
\label{ecltab}
\tablewidth{0pt}
\tabletypesize{\scriptsize}
\tablecaption{Eclipsing Stars}
\tablehead{\colhead{} & \colhead{EPIC ID} & \colhead{BN04} & \colhead{$G$} & \colhead{$(G_{BP}-G_{RP})$} & \colhead{$P$ (d)} & \colhead{$t_0-2450000$} & \colhead{$\mu$} & \colhead{$\bar{\omega}$} & \colhead{Notes}}
\startdata
    & 246887733 &  390 & 12.010 & 0.789 & 8.042 & 7826.97 & M & M & EW? above TO\\
    & 246849982 & 1126 & 12.461 & 0.251 & $>70$ & 7894.17 & NM & NM & EA; 0.1\% variations out of eclipse\\
V4  & 246899376 & 7615 & 12.484 & 0.684 & 2.208595 & 2639.61 & M & M & EA; eccentric; SB2\\
    & 246903724 & 7376 & 12.965 & 0.571 & 2.182 & 7822.81 & M & M & EW?\\
    & 246906509 & 7679 & 13.507 & 0.570 & 2.3185 & 7826.71 & M & M & EB\\
V18 & 246886747 & 7145 & 14.087 & 0.739 & 7.78275 & 2638.2 & M & M & EA\\
    & 246865365 & 4915 & 14.894 & 0.934 & 3.3858 & 7860.4720 & NM? & & exoplanet\\
    & 246899433 &  230 & 15.551 & 1.144 & 13.1076 & 7834.15 & NM & & EA\\
    & 246898423 & 3737 & 15.725 & 0.971 & 7.1277 & 7822.94 & NM & M & EA; no secondary eclipse? or 2P?\\
    & 246860261 &      & 16.307 & 0.820 & 13.107 & 7900.1 & NM & NM? & EA; eccentric; EPIC 246860357 aperture\\
    & 246846293 & 1818 & 16.289 & 1.121 & 8.577 & 7821.74 & NM & NM & EA, weak 2nd eclipse\\
    & 246896535 &      & 17.691 & 0.993 & 46.238 & 7858.853 & M & NM? & EA; CMD nm; EPIC 246896312 aperture\\
V16 & 246901174 &      & 18.722 & 1.421 & 0.2764053 & 2633.355 & NM &  & EW \\
\enddata
\end{deluxetable*}

V1178 Tau (also known as V4; see Figure \ref{ebphase}) is a detached double-lined spectroscopic
binary that contains at least one pulsating star, and its analysis
will be the subject of a separate paper (Hedlund et al., in prep). For
the moment, we report that the orbital period is quite short
(2.208595 d), and despite the strong tidal effects and orbital
circularization that should result from that, the secondary eclipse
does not fall at an orbital phase of 0.5. Further discussion of the
origins of the binary is delayed to the future paper, but we briefly
discuss its pulsations in section \ref{puls}.

One of the new long-period detections (EPIC 246849982; see Figure \ref{singlee}) has an eclipse
of long duration (about 1.2 d) centered on BJD 2457894.17.  The large
drop in flux (about 25\%) and hint of a period of totality is
encouragement for follow-up observations. However, all indications are
that it is not a cluster member: the system photometry puts it
significantly bluer than the cluster turnoff (see Figure \ref{pulscmd}), and {\it Gaia} proper motions
and parallax are well outside the ranges inhabited by cluster stars.

Two objects showed shallow eclipses of exoplanet-like depths.  EPIC
246865365 (see Figure \ref{ebphase}) is a short period system (3.3858
d) with eclipses of 0.5\% depth. \citet{rampalli} recently discussed
this system and identified it as harboring a Saturn-size
exoplanet. {\it Gaia} measurements of this object have large
uncertainties in Data Release 2, but according to the proper motion
membership studies \citep{bnpm,kkpm} the star was identified as a
cluster member.  The photometry of the system is also consistent with
being within the binary star sequence in the cluster
CMD. \citeauthor{rampalli}, however, find that the proper motion and radial velocity of
EPIC 246865365 deviates substantially from the cluster mean, 
and the star does not show variations consistent with having a stellar
companion. Their conclusion is that it is a foreground object. A
  couple of small transit-like events were seen in the light curve of
  EPIC 246896312, but these were traced to a fainter eclipsing binary
  (EPIC 246896535) in the aperture. This appears to be a field star
  based on {\it Gaia} parallax and CMD position.

EPIC 246887733 (see Figure \ref{ewphase}) is a bright system above the
cluster turnoff that has a relatively long period ($\sim8$ d) for the
continual light curve variations that it shows. The object is likely
to contain a subgiant star in weak thermal contact with its companion
based on the difference in eclipse depths. The system has also been
detected in the FUV and NUV bands by {\it GALEX} \citep{galexcat}. Further observations
would be necessary to determine the degree to which the low amplitude of the
light curve comes from a weakly ellipsoidal star or a nearly face-on orbit.

\subsection{Pulsating Stars in the Instability Strip}\label{puls}

Prior to Campaign 13, \citet{frandsen}, \citet{arentoft1817} and
\citet{ander1817} identified 23 main sequence pulsating variables in the field of
NGC 1817. The age of NGC 1817 is such that the cluster turnoff falls
completely in the instability strip, similar to NGC 6811 in the
main {\it Kepler} field \citep{sand6811}. As a result, pulsating
variables are among the brightest stars in the cluster, and their
oscillation spectra may reveal information about the internal changes
that result from their evolution away from the main sequence.

The fields surveyed by previous ground-based studies were considerably
smaller than the field observed by K2, and the precision of the K2
photometry is considerably better, so we expected a large group of new
detections. The identification of frequencies was challenging in the
ground-based datasets due to the complicated structure of the
window function for the observations. In addition, the ground-based
studies put their light curves through a high-pass filter that may have
selected against some modes of oscillation like $g$ modes that are
observed in many main-sequence pulsating stars. For the
purposes of this paper, we will focus on pulsating stars that are
members of the NGC 1817 cluster. In part this is because the original
target lists were selected using ground-based proper motion
information that has since been superseded by {\it Gaia} data. After
{\it Gaia} DR2, we identified all targets observed in K2
Campaign 13 that had high memberhip probability based on proper
motions and parallaxes. So it is quite possible that there are
additional pulsating variables in the field, but the list of pulsating
cluster members that we have compiled is not likely to be complete due
to the selection process.

Strongly outlying datapoints in the time series photometry were
removed before any analysis was conducted.  We used Period04 software
\citep{period04} to compute the pulsation spectra of observed cluster
members with $G < 15$ for frequencies less than the Nyquist frequency
($f \approx 24.5$ day$^{-1}$ or 283 $\mu$Hz) in the long cadence
data. In some cases (especially among the fainter $\delta$ Sct stars),
there are probably oscillation modes above the Nyquist frequency due
to the higher stellar densities. We pre-whitened the spectrum by
successively fitting and subtracting sine waves having the
  frequency with the highest power remaining in the spectrum. Table
\ref{pulstab} contains the frequency and amplitude of the two
strongest frequencies in the spectrum of each star. We generally
identified stars as $\delta$ Sct or $\gamma$ Dor stars based on
whether the strongest frequencies were found above or below 5
day$^{-1}$, which is generally identified as a high frequency
  limit for $g$ modes. Some stars showed frequencies of similar
amplitude on both sides of this boundary, and we classified these as
hybrid stars.
In total, we identify 32 $\delta$ Sct pulsators (20 of them new),
27 $\gamma$ Dor candidates (all but 2 new),
and 7 hybrids (4 new).


\startlongtable
\begin{deluxetable*}{lcrcc|rcccccl}
\tablewidth{0pt}
\tabletypesize{\scriptsize}
\label{pulstab}
\tablecaption{Main Sequence Pulsating Stars}
\tablehead{\colhead{} & \colhead{EPIC ID} & \colhead{BN04} & \colhead{$G$} & $(G_{BP}-G_{RP})$ & \colhead{$f_1$} & \colhead{$A_1$} & \colhead{$f_2$} & \colhead{$A_2$} & \multicolumn{2}{c}{Membership}  & \colhead{Notes}\\ \colhead{} & \colhead{} & \colhead{} & \colhead{} & & \colhead{(day$^{-1}$)} & \colhead{(mmag)} & \colhead{(day$^{-1}$)} & \colhead{(mmag)} & \colhead{$\mu$} & \colhead{$\bar{\omega}$} & \colhead{}}
\startdata
V1  & 246893961 &  167 & 13.413 & 0.625 & 19.9255 & 1.875 & 11.2399 & 1.386 & M & M & $\delta$ Sct; SB\\
V2  & 246894443 &  154 & 12.776 & 0.669 &  0.6082 & 1.259 & 18.9008 & 0.640 & M & M & hybrid\\
V3  & 246895317 &  184 & 14.263 & 0.735 & 18.4952 & 0.216 & 18.8231 & 0.212 & M & M & $\delta$ Sct\\
V4  & 246899376 & 7615 & 12.484 & 0.684 & 11.1044 & 2.180 &  1.4887 & 2.087 & M & M & hybrid?/eclipsing binary\\
V5  & 246888761 &  155 & 12.803 & 0.690 & 14.9202 & 2.377 & 10.9488 & 0.764 & M & M & $\delta$ Sct\\
V6  & 246894145 &  156 & 12.813 & 0.709 & 18.0807 & 0.691 & 15.7912 & 0.449 & M & M & $\delta$ Sct\\
V7  & 246901593 & 7632 & 13.620 & 0.619 & 14.6041 & 1.316 &  9.4671 & 1.313 & M & M & $\delta$ Sct\\
V8  & 246896788 &  423 & 14.205 & 0.743 & 20.1215 & 0.612 & 20.4607 & 0.589 & M & M & $\delta$ Sct\\
V9  & 246897103 & 7294 & 13.074 & 0.785 &  8.8691 & 0.601 & 10.0776 & 0.469 & M & M & $\delta$ Sct; triplet; $f_3 = 7.7095$\\
V10 & 246888938 &  240 & 16.455 & 0.796 & & & & & NM & NM & $\delta$ Sct; below MS\\
V11 & 246893551 & 7097 & 14.192 & 0.705 & 20.5668 & 0.624 & 16.8276 & 0.167 & M & M & $\delta$ Sct; dominant freq\\
V12 & 246891220 & 7105 & 14.533 & 0.799 &  2.5487 & 4.011 &  2.4949 & 3.878 & M? & M & $\gamma$ Dor \\
V13 & 246895148 &  192 & 14.559 & 0.782 & 18.8236 & 0.659 & 18.4952 & 0.585 & M & M & hybrid; gaussian packet\\
V19 & 246882381 &  162 & 13.243 & 0.578 & 11.4611 & 1.121 & 10.6118 & 1.042 & NM & M & $\delta$ Sct; 2 strongest freqs.\\
V20 & 246882504 &  181 & 14.225 & 0.690 & 24.0070 & 0.330 & 18.9431 & 0.126 & M & M & $\delta$ Sct\\
V21 & 246889963 & 7214 & 14.308 & 0.707 & 19.6082 & 0.936 &  9.6786 & 0.672 & M & M & $\delta$ Sct; prob. hits Nyquist\\
V22 & 246905680 & 7332 & 13.854 & 0.703 & 14.1882 & 3.126 & 16.4024 & 0.332 & M & M & $\delta$ Sct; dominant freq.\\%
V23 & 246887060 & 7148 & 14.560 & 0.746 &  1.6020 &10.431 &  1.6415 & 7.948 & M & M & $\gamma$ Dor\\
V26 & 246895228 & 7469 & 16.278 & 1.066 &         &       &         &       & M & M & not variable\\
V1177 & 246897707 & 7298 & 14.017 & 0.636 &       &      &         &       & M & M & not variable\\
& 246851341 & 1180 & 15.342 & 1.021 &  0.7658 & 0.593 &  0.7427 & 0.278 & M? & M & $\gamma$ Dor?\\
& 246855623 &      & 13.405 & 0.677 & 17.9151 & 3.337 & 15.7139 & 0.796 & M & M & $\delta$ Sct; dominant freq.\\
& 246857175 & 1084 & 14.390 & 0.759 & 19.5160 & 1.451 & 19.0023 & 0.949 & M & M & hybrid\\
& 246860219 &      & 12.440 & 0.766 &  0.7946 & 0.197 &  0.8329 & 0.047 & M & M & rotational?\\
& 246861917 &  781 & 13.962 & 0.714 & 13.6109 & 2.461 & 15.6916 & 1.177 & M & M & $\delta$ Sct\\
& 246866279 & 4909 & 14.709 & 0.817 &  2.0092 & 2.175 &  2.1707 & 1.761 & M & M? & $\gamma$ Dor; 2 freq. groups\\
& 246871279 &  775 & 13.083 & 0.706 &  1.3957 & 0.130 &  1.4799 & 0.063 & M & M & rotational?\\
& 246874980 & 4907 & 14.592 & 0.773 &  1.5549 & 3.370 &  1.8342 & 2.250 & M & M & $\gamma$ Dor\\%
& 246878285 &   51 & 14.766 & 0.895 &  0.9422 & 0.192 &  0.9617 & 0.131 & M & M & $\gamma$ Dor? near single freq.\\
& 246888228 & 7085 & 15.430 & 0.916 &  0.7688 & 0.104 &  3.8347 & 0.076 & M & M & $\gamma$ Dor?\\
& 246883000 &  160 & 13.024 & 0.610 &  1.2971 & 0.326 &  1.2110 & 0.153 & M & M & hot $\gamma$ Dor?\\
& 246884149 & 7746 & 12.198 & 0.696 &  0.8203 & 0.048 &         &       & M & M & rotational?\\
& 246884792 &  889 & 14.109 & 0.717 & 18.4841 & 1.110 &  1.3012 & 1.110 & M? & M & $\delta$ Sct\\
& 246886811 & 7753 & 14.685 & 0.789 &  1.5471 & 1.536 &  1.5378 & 0.491 & M & M & $\gamma$ Dor\\
& 246887076 & 7768 & 13.043 & 0.745 &  6.3433 & 2.171 &  9.4438 & 0.292 & M & M & $\delta$ Sct; dominant freq.\\
& 246887105 &  525 & 14.287 & 0.711 &  0.6796 & 0.423 & 20.0530 & 0.392 & M & M & hybrid; weak; gaussian env.\\%
& 246888602 & 7154 & 13.092 & 0.573 & 11.3395 & 0.321 &         &       & M & M & $\delta$ Sct; dominant freq\\%
& 246888903 & 3683 & 13.913 & 0.649 & 17.5103 & 0.456 & 21.7653 & 0.307 & M & M & $\delta$ Sct; prob hits Nyquist\\
& 246889429 & 6660 & 14.122 & 0.636 & 22.8715 & 2.597 & 22.6384 & 1.798 & M? & M? & $\delta$ Sct; beats; prob hits Nyquist\\
& 246890269 & 3698 & 14.571 & 0.756 &  1.8021 & 2.011 &  1.9146 & 1.576 & M & M & $\gamma$ Dor\\
& 246890414 &  163 & 13.203 & 0.681 & 15.2372 & 0.189 & 15.9872 & 0.161 & M & M & $\delta$ Sct\\
& 246891489 & 7078 & 13.776 & 0.648 & 15.8220 & 0.549 &         &       & M & M & $\delta$ Sct; dominant freq.\\
& 246892400 & 7079 & 12.754 & 0.763 &  9.5211 & 0.371 &  8.4558 & 0.357 & M & M & $\delta$ Sct; beats\\
& 246892423 & 7225 & 14.468 & 0.783 &  2.2597 & 1.016 &  1.3459 & 0.975 & M & M & $\gamma$ Dor\\%
& 246892620 & 3671 & 13.134 & 0.548 & 22.1865 & 0.319 & 22.2267 & 0.298 & M & M & $\delta$ Sct\\
& 246892892 &  150 & 12.691 & 0.646 &  2.5376 & 0.120 &  1.3084 & 0.108 & M & M & hot $\gamma$ Dor? \\
& 246893663 &  196 & 14.662 & 0.870 &  0.9843 &11.434 &  0.8780 & 9.558 & M? & M? & $\gamma$ Dor\\
& 246895037 &  202 & 14.799 & 0.816 &  1.9089 & 0.133 &  1.9403 & 0.114 & M & M & $\gamma$ Dor\\
& 246895872 & 3674 & 13.343 & 0.749 & 14.8691 & 0.314 & 23.6892 & 0.128 & M & M & $\delta$ Sct; dominant freq.\\
& 246895958 &  195 & 14.625 & 0.792 &  2.9783 & 1.561 &  2.7832 & 1.396 & M & M & $\gamma$ Dor\\
& 246896275 & 7467 & 12.900 & 0.312 &  1.5452 & 0.107 &  1.8216 & 0.096 & M & M & rotational?\\
& 246897854 & 3670 & 12.953 & 0.617 &  1.1061 & 0.183 &  3.8177 & 0.170 & M & M & hybrid? \\
& 246898842 & 6684 & 15.199 & 0.886 &  0.9064 & 0.306 &         &       & M & M & $\gamma$ Dor?\\
& 246899036 & 171  & 13.639 & 0.640 &  2.2323 & 0.186 & 12.9797 & 0.114 & M? & M? & hybrid \\
& 246899163 & 7362 & 13.712 & 0.645 & 14.9972 & 0.132 & 13.4081 & 0.113 & M & M & $\delta$ Sct\\
& 246899781 & 6669 & 14.341 & 0.784 &  4.9691 & 1.905 &  4.9150 & 1.358 & M & M & $\gamma$ Dor; strong oscill.\\
& 246899840 &  186 & 14.165 & 0.833 &  1.3541 & 2.808 &  1.3869 & 1.655 & M & M & $\gamma$ Dor\\
& 246902309 & 7308 & 13.006 & 0.729 &  9.4269 & 0.531 & 13.0180 & 0.381 & M & M & $\delta$ Sct\\
& 246902508 & 7641 & 13.133 & 0.645 & 14.4388 & 0.366 & 17.3904 & 0.367 & M & M & $\delta$ Sct\\
& 246904294 & 7311 & 14.328 & 0.729 &  7.4606 & 1.731 &  2.1215 & 0.998 & M & M & hybrid; 3 rich packets\\%
& 246905414 &      & 14.496 & 0.686 &  2.0456 & 3.268 &  1.8529 & 1.500 & M & M & $\gamma$ Dor; beats\\
& 246912283 &      & 13.037 & 0.585 & 18.5666 & 0.766 & 17.9832 & 0.546 & M & M & $\delta$ Sct\\
& 246912694 &      & 13.721 & 0.605 & 14.0664 & 0.647 & 19.0229 & 0.326 & M & M & $\delta$ Sct\\
& 246914712 &      & 13.440 & 0.620 &  9.6573 & 0.240 & 13.5149 & 0.194 & M & M & $\delta$ Sct\\
& 246916477 &      & 12.935 & 0.627 & 15.6072 & 4.272 & 17.4394 & 4.046 & M & M & $\delta$ Sct; beats\\
& 246919500 &      & 14.438 & 0.685 &  1.9357 & 5.592 &  1.7059 & 3.351 & M & M & $\gamma$ Dor\\
& 246919843 &      & 11.950 & 0.956 & 15.8546 & 0.218 & 17.4526 & 0.129 & M & M & $\delta$ Sct; subgiant?\\
& 246920074 &      & 14.428 & 0.727 &  2.4856 & 2.871 &  2.6321 & 1.997 & M & M & $\gamma$ Dor, beats\\
& 246922458 &      & 14.558 & 0.746 &  1.6315 & 3.631 &  1.0517 & 2.849 & M & M & $\gamma$ Dor\\
\enddata
\end{deluxetable*}


\begin{figure}
  \epsscale{1.3}
  \plotone{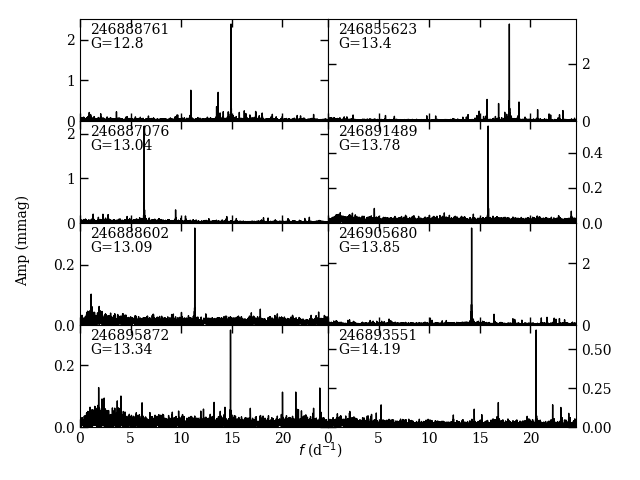}
\caption{Amplitude spectra for main sequence pulsating stars with
  a single high amplitude mode in NGC 1817.
\label{periodo}}
\end{figure}

Figure \ref{pulscmd} shows the CMD positions of the detected pulsating
stars. It is clear that many of the brighter stars have evolved away
from the main sequence, and are likely to have lower average densities
than the fainter objects on the main sequence. We expect this to
affect the fundamental mode of oscillation, and generally speaking,
the average frequency of oscillations that are excited.  However, even
among stars with a single dominant mode peak (see Figure
\ref{periodo}) there is not a clearly visible relationship between the
frequency and the luminosity of the star (which should relate to its
average density).  This is just one example of the difficulty of
identifying the mode of oscillation in main sequence pulsating stars,
and we will discuss this more below.

\subsubsection{Ensemble Identification of Pulsation Modes}

\citet{arentoft1817} looked at the strongest modes of oscillation as a
function of magnitude for the ensemble of $\delta$ Sct stars they
detected in NGC 1817. They did this in the hope that the relative
magnitudes of the stars could be related to differences in their
average densities, which should affect the frequency of a particular
mode in a predictable way, allowing modes to be identified. In
practice, mode frequencies can be shifted by complicating factors like rotation and binarity,
and in the case of NGC 1817, the evolutionary state of different
stars.  In an examination of a cluster of similar age (NGC 6811),
\citet{sand6811} found evidence of a period-luminosity relation among
the largest amplitude modes observed in the $\delta$ Sct
stars. \citet{ziaali} recently found similar evidence among $\delta$
Sct stars observed with {\it Kepler} and with luminosities derived
from {\it Gaia} DR2 parallaxes. The radial fundamental mode of
oscillation is likely to be among the few modes at the low frequency
end of the spectrum, and this feature may help set up a P-L relation
for these stars.

\begin{figure*}
\plotone{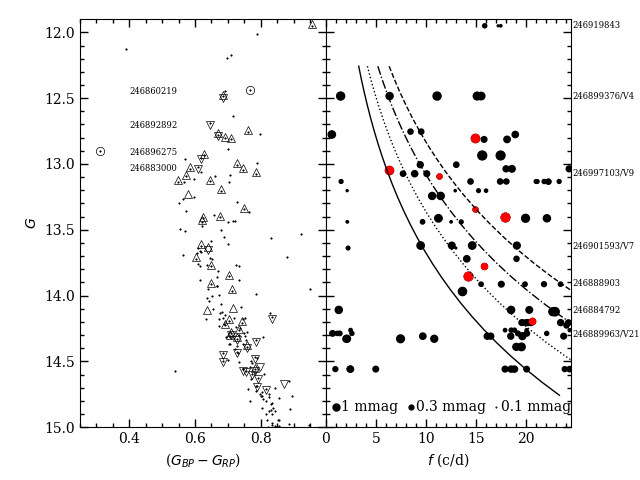}
\caption{{\it Left:} {\it Gaia} color-magnitude diagram positions for
  main sequence pulsators in NGC 1817. $\delta$ Sct stars are shown
  with $\triangle$, $\gamma$ Dor stars with $\triangledown$, hybrid
  stars with a star symbol, and possible rotating variables with
  $\bigcirc$. EPIC IDs for the bright low-frequency variables are
  shown at the same magnitude level. {\it Right:} Strongest
  frequencies (any mode with amplitude greater than half that of the
  one with largest amplitude) versus magnitude for stars with
  high-frequency pressure modes of oscillation ($\delta$ Sct or hybrid
  pulsators). Symbol size is related to the amplitude of the mode, and
  red symbols show stars with a single dominant mode of
  pulsation. Lines (from left to right) show predicted positions of
  the radial fundamental mode and first through third overtones using
  the \citet{mcnamara} P-L relation and \citet{stelling} period
  ratios, along with a cluster distance modulus $(m-M)_G = 12.06$.
  EPIC IDs of selected variables are shown at the magnitude level of
  the star's oscillation modes.
  \label{plrel}}
\end{figure*}

\begin{figure}
  \epsscale{1.3}
  \plotone{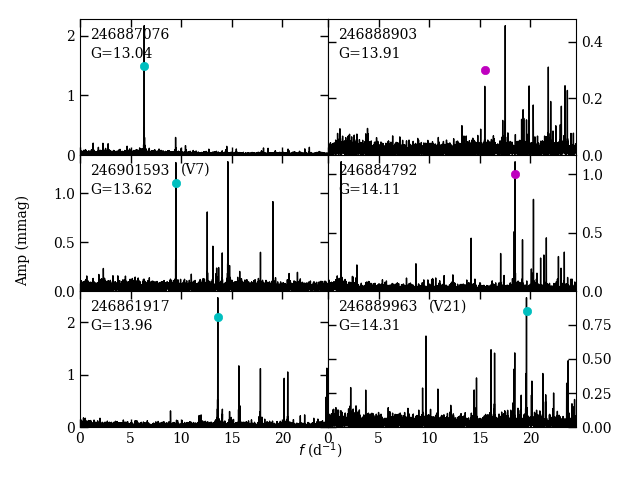}
\caption{Amplitude spectra for stars with possible strong radial fundamental
  mode (cyan marks; left panels and bottom right panel) and first
  overtone (magenta marks; top two right panels) pulsation in NGC 1817.
\label{fundament}}
\end{figure}

Figure \ref{plrel} shows the zoomed color-magnitude diagram of the
pulsators, and a diagram of $G$ magnitude versus frequency for the
strongest modes. When we shift the period-luminosity relation
\citep{mcnamara} for the fundamental mode oscillations of
high-amplitude $\delta$ Sct stars with Hipparcos parallaxes
(solid curve in the right panel) using a distance modulus based on
the {\it Gaia} mean parallax and an extinction $A_G$ based on the
reddening $E(B-V) = 0.28$, we find excellent agreement with the low
frequency envelope of the strongest observed modes. This envelope is
set by a handful of stars: EPIC 246887076 (a star having a single
dominant frequency), EPIC 246901593/V7, and possibly EPIC 246861917
and EPIC 246889963/V21. \citet{arentoft1817} identified three stars in
their ground-based data that had strongly excited modes near the
fundamental, but only one of these (V7) is in our list above.  We find
that the other two (V1 and V9) have frequencies close to the
low-frequency envelope, but significantly above. V9 is discussed in
subsection \ref{v9}. Figure \ref{fundament} shows the spectra of the
most likely fundamental mode pulsators.

Theoretical models indicate that period ratios for radial
  pulsation modes are stable \citet{stelling}, so if we have reliably
  identified the frequency of the fundamental mode as a function of
  magnitude, we should be able to predict the frequencies of
  overtones. The dotted and dashed curves in Figure \ref{plrel} show
  predicted values for first through third overtones using theoretical
  period ratios. Once again, rotation, binarity, and evolutionary
  effects shift frequencies and will complicate identification for
  most stars, but we looked for ones with the strongest cases.  The
variable V7 may be the most interesting of the stars because its
fourth strongest frequency agrees well with the predicted frequency of
the first overtone if its second strongest frequency is indeed the
fundamental mode. In addition, the strongest mode falls near the
predicted second overtone, and the third strongest mode falls near the
third overtone.

There are a few stars that have patterns in their frequency spectra
that may identify a radial overtone mode. EPIC 246888903 (see Figure
\ref{fundament}) shows 5 strong frequencies that appear to be a good
match for the first, second, and third radial overtones,
along with two other modes that sit approximately halfway
between the overtones. EPIC 246884792 shows a strong mode that matches
the approximate frequency predicted for the first radial overtone.

Other stars may be fundamental mode pulsators, but
binary companions could affect their magnitudes and shift them
vertically away from the low-frequency envelope. An example of this is
the bright pulsating star EPIC 246899376 (V4 / V1178 Tau), which
contains two stars of nearly equal brightness, and is discussed
below.

\subsubsection{The Eclipsing System V1178 Tau}

V1178 Tau (EPIC 246899376) was first identified as an eclipsing binary
star in \citet{arentoft1817}. They gave the star the identifier V4,
and they detected 4 $\delta$ Sct pulsation frequencies. \citet{mz1817}
found that the binary is double-lined.

\begin{figure*}
\plotone{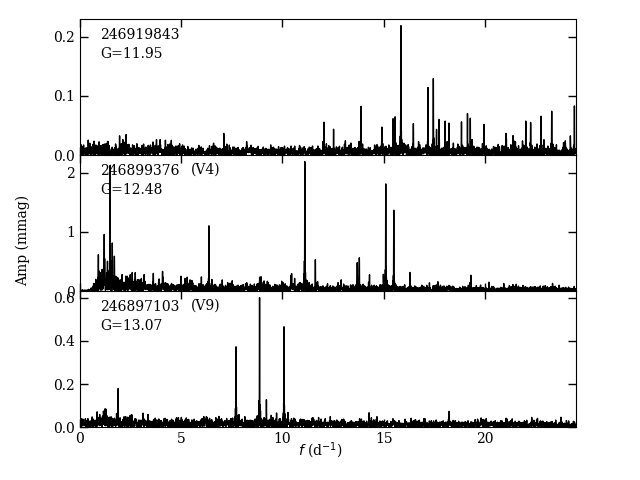}
\caption{Amplitude spectra for pulsating stars of extra interest 
in NGC 1817.
\label{interest}}
\end{figure*}

We derived a clean pulsation spectrum by subtracting an optimized
model eclipsing light curve using the program ELC \citep{elc}. For our
purposes in this paper, we are only interested in the pulsation
spectrum, and so we will not discuss the parameters used in the
fit. The spectrum derived from Period04 is shown in Figure \ref{interest}. There
are five frequencies with amplitudes of more than 1 mmag in the K2
data set. Three of the detected frequencies roughly correspond to the
detections by \citeauthor{arentoft1817}, but our second highest
amplitude detection is in the low frequency $g$ mode range at 1.488 day$^{-1}$. This mode is the main reason we label the system a possible
hybrid. In addition, we do not see any sign of a mode at 8.32 day$^{-1}$ as
reported by \citeauthor{arentoft1817}.

High-resolution spectra show lines of both stars, and for spectra of
stars with similar temperatures, broadening function
\citep{rucinski02} areas can be used as proxies for luminosity.
  Therefore we conducted a preliminary analysis of HARPS spectra to
  derive broadening functions for the two stars. After fitting the
  broadening functions with Gaussians, we derived an area ratio of
  0.85. If this represents the luminosity ratio properly, the
pulsating star should be between 0.67 and 0.84 mag below the magnitude
of the binary.  If this is taken into consideration, the fifth
strongest detected mode with frequency 6.368 day$^{-1}$ falls near the
radial fundamental mode line in Figure \ref{plrel}. This possibility
makes the system worth additional study. In a binary with a relatively
short orbital period of around 2.2 days, the orbit should have
circularized and the stellar rotation should have synchronized with
the orbit for an age near 1 Gyr. However, the orbit has significant
eccentricity based on the eclipse spacings in phase (see section
\ref{ecl}), which is a clue that the system might have been involved
in dynamical interactions after the cluster's formation.  Another
consequence of this is that the rotation of the stars may still be
slowing down or may be pseudo-synchronized with the orbital
  motion at periastron.  Preliminary measurements of $v \sin i$ from
the broadening functions indicate speeds near 70 km s$^{-1}$, which is
much slower than typical main sequence A stars. The rotations of
  both stars are faster than they would be if synchronized ($\sim50$
  km s$^{-1}$) with the orbital motion, but the primary star may be
  pseudo-synchronized (using $e approx 0.13$ and $R_1 \approx
  2.2\rsun$; \citealt{hut}). While there are large samples of
  eclipsing binaries with $\delta$ Sct pulsators now known
  \citet{liakos,kahraman,gg}, short period binaries with eccentric
  orbits are rare. We find three detached systems that are identified
  as eccentric in the tabulation of \citeauthor{liakos}, but in all
  cases the eccentricity is marginal. Because V1178 Tau is a member of
  the NGC 1817 cluster, the age may provide some interesting
  constraints on the dynamics of the system for future studies.

\subsubsection{A Mode Triplet?}\label{v9}

EPIC 246897103 (V9) was identified by \citet{arentoft1817} as a
$\delta$ Sct star from the ground-based observations, and they
identified three pulsation frequencies. In the amplitude spectrum of
the K2 light curve, we find four strong modes, but only one is a match
to the three in \citeauthor{arentoft1817}, probably due to the complex
window function for the ground-based observations. The star is of
interest because the three strongest frequencies (8.869, 10.077, and
7.710 day$^{-1}$) resemble a mode triplet that may be related to
strong rotation (see Figure \ref{interest}). The star is the reddest
pulsating star at the extended cluster turnoff, consistent with the
idea that strong stellar rotation may be responsible for the color
spread in intermediate-age clusters like NGC 1817 and NGC 5822
\citep{sun5822}. \citet{mz1817} made a low-resolution spectroscopic
measurement of the rotational velocity, and found $v \sin i = 220$ km
s$^{-1}$. We can estimate the rotational frequency as $\nu_{rot}
  = v_{rot} / 2\pi R \approx 2$ day$^{-1}$, assuming a stellar radius
  of about $2.2\rsun$ similar to the primary star of V1178
  Tau. Because the rotational frequency spacing is likely a multiple
  of $\Omega_{rot} = 2\pi \nu_{rot}$ and more than a factor of 10
  larger than the frequency spacing of the modes above, rotation may
  not be a good explanation unless there is a serious error in one of
  the measurements. Effects due to a highly inclined spin axis are not
  plausible here due to the large measured rotational velocity.
Alternately, we note that the second and third strongest frequencies
fall near the predicted frequencies of the second and first radial
overtones, respectively (see Figure \ref{plrel}.)

\subsubsection{A Subgiant Candidate}

Weak variability is detectable in the bright star EPIC 246919843.
  Because it sits in a very sparsely populated part of the cluster's
  CMD between the main sequence turnoff and the red clump, the star's
  membership should be validated.  However, all of the information
  available (including its {\it Gaia} parallax and radial velocity)
  indicates that it is a high probability cluster member.  The CMD
position of the object gives it a chance of being either a binary
composed of a faint giant star and a bright main sequence star, or a
single subgiant star. If the object is a binary, both stars could
potentially be variable (an asteroseismic giant and a pulsating main
sequence star, for example), so we looked at the spectral energy
distribution (SED) first (see \S \ref{seddat}).

\begin{figure}
\epsscale{1.3}
    \plotone{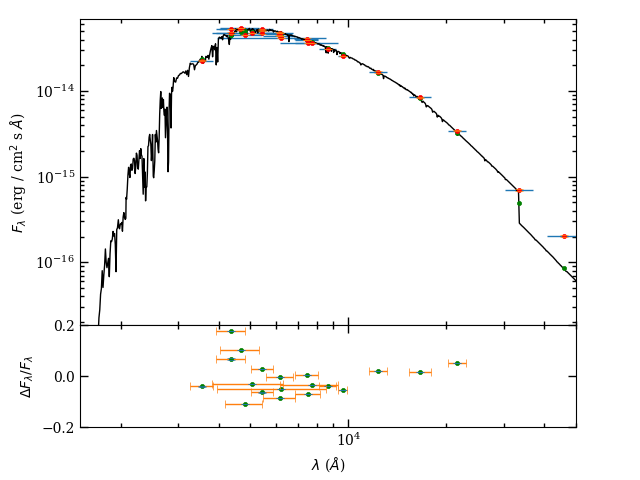}
\caption{{\it Top panel:} Spectral energy distribution for the star
  246919843 from photometry (orange points), compared with an ATLAS9
  model with $T_{\rm eff} = 6250$ K, $\log g = 4.0$, and $E(B-V) =
  0.23$. Horizontal bars show the approximate wavelength range of each
  filter used. {\it Bottom panel:} Fractional flux residual between
  the observations and the model fit.
\label{subgi}}
\end{figure}

For comparison, we used theoretical models from ATLAS9 \citep{atlas9} with
adjustment of the model to account for an interstellar reddening of
$E(B-V)=0.23$ using the \citet{cardelli} extinction curve. Model
photometry was calculated using the IRAF routine {\tt sbands}.  The
SED is well fit with a temperature of about 6250 K (see
Fig. \ref{subgi}), appropriate for a star between the main sequence
and red clump. For comparison, a sample of NGC 1817 $\delta$ Sct stars were found
by \citet{mz1817} to have effective temperatures between about 6900
and 8100 K.

The relatively large amplitude and high frequency of the variations argue against the
possibility of a binary containing a red clump star --- the detected
frequencies are centered at around 16 day$^{-1}$ or 184 $\mu$Hz, larger than
any of the giants detected in NGC 1817 or the similar cluster NGC 6811
(see Figure \ref{interest}).  Similarly, while the oscillation
frequencies lie in the range populated by $\delta$ Sct pulsation
modes, they appear at too high a frequency for the relatively low
density of a subgiant star's envelope unless they are higher
overtones.
Our conclusion is that this is a genuine subgiant star in the cluster
within the Hertzprung gap. Although stars like this are rare, 
it is most probable to find one in a heavily populated
cluster like NGC 1817.

\subsubsection{Unusual $\gamma$ Dor-Like Stars}

\begin{figure}
\epsscale{1.3}
  \plotone{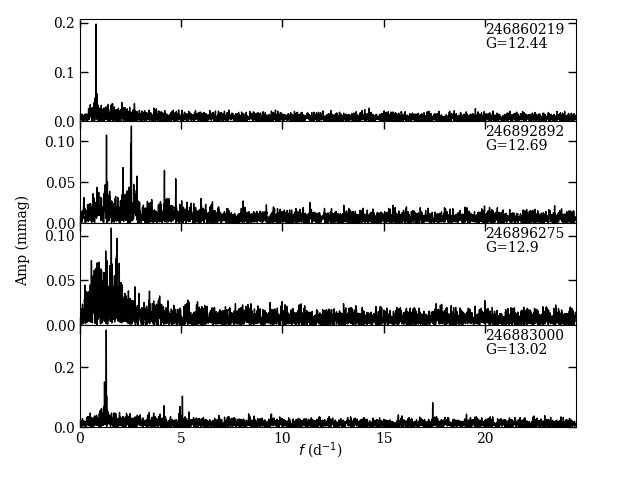}
\caption{Amplitude spectra for  stars with $\gamma$ Dor-like pulsation in NGC 1817.
\label{gdorlike}}
\end{figure}

The instability strip for $\gamma$ Dor stars is observed to overlap
with the cool edge of the $\delta$ Sct instability strip for field
stars (e.g. \citealt{balona18}), and the majority of the NGC 1817 $\gamma$ Dor
stars we identified show this as well (see the left
  panel of Figure \ref{plrel}). However, we also discovered a handful
  of much brighter (and sometimes hotter) stars with $\gamma$ Dor-like pulsation. The
  pulsation spectra of these stars are shown in Figure \ref{gdorlike}.

EPIC 246896275 is the bluest variable star identified in this study,
and as with all of the stars in this section, {\it Gaia} parallax and
proper motion information is completely consistent with cluster
membership. This makes EPIC 246896275 a likely blue straggler. The low
frequency spectrum appears to have a roughly Gaussian-shaped
  envelope, similar to rotational candidates in
\citet{balona13}. However, such features have not been explained yet.

EPIC 246892892 is brighter than the cluster turnoff, and if it is a
single star, it would have to have started its evolution off of the
main sequence. We find at least seven low amplitude frequencies
between 1.30 and 4.72 day$^{-1}$, with four between 2.12 and 2.81
day$^{-1}$. The indications are that these are real pulsation modes,
which would also make this a possible hot $\gamma$ Dor star \citep{balona}.

EPIC 246860219 is one of the brightest main sequence stars identified
with pulsations, and its spectrum is dominated by a single frequency
(0.795 day$^{-1}$). We tentatively identify the star as a rotational
variable candidate, similar to stars discussed in \citet{balona13}. If
true, the detected frequency should be related to the rotation period
of the star.

EPIC 246883000 is found among the $\delta$ Sct pulsating stars at the
cluster turnoff. Its spectrum is also dominated by a single frequency
(1.297 day$^{-1}$), but there are several other weaker low frequencies (1.211,
5.058, and 4.158 day$^{-1}$) and one higher frequency (17.422 day$^{-1}$)
detected. Based on the pulsation modes and the CMD position, we
classify the star as a hot $\gamma$ Dor candidate.

\section{Conclusions}

We examined the K2 light curves for hundreds of stars in the
massive open cluster NGC 1817. This dataset gives us the opportunity
to examine large uniform samples of stars in different evolutionary
phases within this approximately 1 Gyr old, slightly subsolar
metallicity population.

Photometric and asteroseismic evidence shows that the cluster is
producing a large fraction of its red clump stars with nearly the
minimum helium core mass. NGC 1817 is on the young side of the range
of ages that can produce these stars. More precise measurement of the
masses of these stars could allow a calibration of the amount of
convective core overshooting on the main sequence.

We find two giant stars in relatively short period binary systems that
show photometric variations that may vary in phase with the
orbits. We determine the periods of previously known eclipsing
systems, and detect new systems. We identify one (V1178 Tau) that has
an eccentric orbit despite a rather short period, which probably
indicates dynamical interactions after the cluster's birth. Another
system (EPIC 246887733) probably contains a subgiant star, and shows
an EW light curve despite an orbital period of more than 8 days.

The turnoff of the cluster contains a rich population of pulsating
variable stars of $\delta$ Sct and $\gamma$ Dor classes. The large
$\delta$ Sct population in particular gives some hope for using
ensemble asteroseismology to help identify pulsation modes in their
rich frequency spectra. The large sample includes unusual stars like a
subgiant star pulsating in high-order modes (EPIC 246919843), and two
hot $\gamma$ Dor candidates (EPIC 246892892 and 246883000).

\acknowledgments This paper includes data collected by the {\it K2}
mission, and we gratefully acknowledge support from the NASA Science
Directorate under grant NNX17AE94G.  D.S. is the recipient of an
Australian Research Council Future Fellowship (project number
FT1400147). Funding for the Stellar Astrophysics Centre is provided by
The Danish National Research Foundation (Grant agreement no.:
DNRF106).

This research made use of the SIMBAD database, operated at CDS, Strasbourg,
France;
the WEBDA database, operated at the
Institute for Astronomy of the University of Vienna;
and the Mikulski Archive for Space Telescopes (MAST). STScI is operated by the
Association of Universities for Research in Astronomy, Inc., under
NASA contract NAS5-26555. Support for MAST is provided by the NASA
Office of Space Science via grant NNX09AF08G and by other grants and
contracts.

\facilities{Kepler, Gaia, GALEX, MAST, CDS}

\software{K2SFF pipeline \citep{k2av},
Period04 \citep{period04}, ELC \citep{elc}}

\end{document}